\documentclass[10pt,aps,prb,twocolumn,superscriptaddress]{revtex4-2}
\usepackage[utf8]{inputenc}
\usepackage{graphics}
\usepackage{epsfig}
\usepackage{dcolumn}% Align table columns on decimal point
\usepackage{bm}% bold math
\usepackage{amsmath} 
\usepackage{dsfont} 
 \usepackage{mathtools}
 \usepackage{amsfonts}
\usepackage{latexsym}
\usepackage{amssymb} 
\usepackage{color} 
\usepackage{hyperref}

 \usepackage{bbm}
\usepackage{bbold}

\usepackage{amsfonts}

 \newcommand{\g}{\bf}
 \usepackage{braket}

\begin{document}
\title{Quantum anomalous Hall insulator   in ionic Rashba lattice of correlated electrons}
  \author{Marcin M. Wysoki\'nski}   \email{wysokinski@magtop.ifpan.edu.pl}
  \affiliation{International Research Centre MagTop, Institute of
    Physics, Polish Academy of Sciences,\\ Aleja Lotnik\'ow 32/46,
    PL-02668 Warsaw, Poland}
    \author{Wojciech Brzezicki}   \email{brzezicki@magtop.ifpan.edu.pl}
     \affiliation{International Research Centre MagTop, Institute of
      Physics, Polish Academy of Sciences,\\ Aleja Lotnik\'ow 32/46,
      PL-02668 Warsaw, Poland}
       \affiliation{Institute of Theoretical Physics, Jagiellonian University,\\
       Prof. Stanis\l{}awa \L{}ojasiewicza 11, PL-30348 Krak\'ow, Poland}

\begin{abstract}
In this work, we propose an exactly solvable two-dimensional lattice model of strongly correlated electrons that realizes a quantum anomalous Hall insulator with Chern number $\mathcal{C}=1$. First, we show that the interplay of ionic potential, Rashba spin-orbit coupling and Zeeman splitting leads to the appearance of quantum anomalous Hall effect. Next, we calculate in an exact manner Chern number for the correlated system where electron-electron interactions are introduced in the spirit of Hatsugai-Kohmoto model using  two complementary methods, one relying on the properties of many-body groundstate and the other utilizing single-particle Green's function, and subsequently we determine stability regions. By leveraging the presence of inversion symmetry we find boundaries between topological and trivial phases on the analytical ground.
Notably, we show that in the presence of correlations onset of topological phase is no longer signalled by a spectral gap closing consistently with phenomenon called in literature as {\it first-order topological transition}. We provide a clear microscopic understanding of this inherently many-body feature by pinpointing that the lowest energy excited states in the correlated system are no longer of the single-particle nature and thus are not captured by a spectral function.   
\end{abstract}  
\maketitle
  \section {Introduction}
Research on the intersection of topological phases and strongly correlated systems has attracted   significant attention in  recent years \cite{Rachel_2018}.
 This is mostly because the description of topological properties of the correlated matter remains an open, though intensively studied \cite{Fisher2011,Assaad2013,Maska2018}, problem, as most available classifications concern non-interacting systems \cite{atland1997,Chiu13,Chiu14,Sato14,Shio16}.

Apart from rare examples, especially concerning phenomenon called {\it first-order topological transitions} \cite{Budich2013,Amaricci_2015,Amaricci_2016,Troyer2016,Roy2016,Barbarino2019},  usually correlations are taken into account together with non-trivial topology by mapping effects of many-body interactions into the effective single-particle picture possibly encompassing symmetry breaking phases. In that manner the resulting effective Hamiltonian is suitable for topological analysis within the known classifications. Illustrative examples not involving symmetry broken states are for instance topological Kondo insulators  where the effect of strong correlations is mostly accounted for by renormalization of hybridization gap and up-shift of the atomic level of $f$-electrons \cite{Coleman_2010,Wysokinski_2016}. On the other hand, a well-known example involving symmetry broken states is the one proposed by Raghu {\it et al} \cite{Raghu_2008} where  appearance of topological state, and also opening of the charge gap is linked to the  onset of interaction driven charge density wave.      

In the present work we address the problem of interplay between correlations and topology from a different angle \cite{Morimoto2016}, which recently gained some attention \cite{Yang2019,arxiv1,arxiv2}. Namely,  we aim on analyzing effect of correlations on a quantum anomalous Hall (QAH) insulator  by taking into account only part of the full Hubbard type of interaction in the Hatsugai-Kohmoto (HK) spirit  \cite{HK}, i.e. interaction term local in momentum space,
that, when strong enough, leads to the opening of the charge Mott gap that can be analyzed with the exact calculations. In the past such a form of interaction has been used for description of so-called statistical spin-liquid \cite{Byczuk1994,Byczuk1995} and its instability towards superconductivity \cite{Spalek1994}. Moreover, more recently, the same interaction was advocated to be essential when it comes to the understanding of discrete symmetry breaking on a Mott metal to insulator transition \cite{Nature1} as well as properties of high $T_c$ cuprate superconductors \cite{Nature2}.

There are several systems realizing or being predicted to realize QAH effect \cite{Chang2013,Chang2015,Sharpe2019,Deng2020,Serlin2020,Satake2020,Fijalkowski2020,Pournaghavi2021,Li2021,Hussain2023} whose topological properties derive from the band structure character, specific form of spin-orbit coupling and magnetic order. Here, we propose yet another  model system realizing QAH as a result of interplay between ionic potential, Rashba spin-orbit interaction and Zeeman splitting,
that up to the best of our knowledge, was not yet reported \cite{Chang2023}.
It is promising for potential experimental realization that mentioned ingredients of the model system can be found altogether either in a synthetic version in the optical lattice with engineered spin-orbit coupling \cite{Meng2016,Huang2016} or in thin layers of ionic insulators doped with magnetic ions.

Our exact analysis of the topological properties of the proposed model in the presence of correlations introduced in HK spirit  performed with two complementary techniques, based on exact many-body groundstate as well as single particle Green's function \cite{Zhang_2012,Zhang_2012_2}  unveiled that (i) the QAH state, though in narrower range of parameters than in uncorrelated case, survives despite high but finite values of interactions and (ii) the spectral gap is not closing at the topological phase transition even for small values of interactions, consistently with the phenomenon called in literature as {\it first-order topological phase transition} \cite{Amaricci_2015,Amaricci_2016,Troyer2016,Roy2016,Barbarino2019}. The latter finding, being a clear manifestation of the inherently many-body nature of the considered system is directly explained on the microscopic ground. It also constitutes another example \cite{Yang2019} where analysis of single-particle Green's function in a correlated system can provide misleading conclusions related to topological properties.

\section{Model}
Our starting point is a single-orbital ionic Rashba model in the presence of Zeeman splitting on the bipartite square lattice 
\begin{equation}
    \begin{split}
    \mathcal{H}_0&=-t\sum_{\langle{\g i}{\g j}\rangle\sigma}c_{\g i\sigma}^\dagger c_{\g j\sigma}+
\sum_{\g i\sigma} [V(-1)^{\g i}+\sigma h-\mu]n_{{\g i}\sigma}\\
&+
\sum_{\langle{\g i}{\g j}\rangle}  \left( i \alpha  (c_{{\g i} \uparrow}^\dagger,c_{{\g i} \downarrow}^\dagger)({\g r_{\g i\g j}} \times \boldsymbol{\sigma})_z\begin{pmatrix}c_{{\g j}, \uparrow}\\c_{{\g j}, \downarrow}
\end{pmatrix} +H.c.\right)
    \end{split}
\end{equation}
where $t$ is the nearest-neighbour hopping, $V$ is the strength of the ionic potential, $(-1)^{{\g i}\in A}=1$ and  $(-1)^{{\g i}\in B}=-1$,  $h$ Zeeman field, $\alpha$ amplitude of Rashba spin-orbit interaction, ${\g r_{\g i\g j}} $ measures the distance between sites $\g i$ and $\g j$,  $\boldsymbol{\sigma} =\{\sigma_x,\sigma_y,\sigma_z\}$ is vector of  Pauli matrices. Hereafter, we set $t$ as an energy unit, i.e. $t=1$. 

After Fourier transformation to the reduced Brillouin zone (RBZ) above model can be cast  into  the following form 
\begin{equation}
\begin{split}
\mathcal{H}_0=
\!\!\!\!\!\sum_{{\g k} \in {\rm RBZ}}\!\!\!\! \hat \psi^\dagger_{\g k }\Big(&\sigma_z\!\otimes\!\big[\Re(g_{\g k})\sigma_x+\Im(g_{\g k})\sigma_y+\epsilon_{\g k} \mathbb{1}\big]\\
&+V\sigma_x\!\otimes\!\mathbb{1}+h\mathbb{1}\!\otimes\!\sigma_z-\mu\mathbb{1}_4\Big) \hat \psi_{\g k }
% -\frac{U}{2}\hat \psi_{\g k }^\dagger \hat \psi_{\g k }
\end{split}
\end{equation}
where ${ g_{\g k}\!=\!2\alpha \sin k_y\!-\!i2\alpha \sin k_x}$, $\epsilon_{\g k}\!=\!-2t(\cos k_x\!+\!\cos k_y)$, $\hat\psi^\dagger_{\g k}=\{c^\dagger_{{\g k},\uparrow},c^\dagger_{{\g k},\downarrow},c^\dagger_{{\g k+Q},\uparrow},c^\dagger_{{\g k+Q},\downarrow}\}$ with $Q=\{\pi,\pi\}$. 
 
 At this stage we enrich single particle Hamiltonian $\mathcal{H}_0$ with   many-body correlations introduced in the Hatsugai-Kohmoto spirit \cite{HK},
 \begin{equation}
  \mathcal{H}=\mathcal{H}_0+\sum_{{\g k} \in {\rm RBZ}}U (n_{\g {k}\uparrow}n_{{\g k}\downarrow}\!+\!n_{\g {k+Q}\uparrow}n_{\g{k+Q}\downarrow})
 \end{equation}
Considered many-body model due to local in momentum form of the interaction is exactly solvable. Throughout the whole paper we set $\mu=U/2$ what ensure  half-filling independently of the other parameters.

As a side remark we note that HK interaction can be seen as a part of the Hubbard interaction,
\begin{equation*}
    U\sum_{\g i} n_{{\g i}\uparrow}n_{{\g i}\downarrow}=\frac{U}{N}\sum_{\g k,p,q}c^\dagger_{{\g k}+{\g p}-{\g q}\uparrow}c^\dagger_{{\g q}\downarrow}c_{{\g p}\uparrow}c_{{\g k}\downarrow}
\end{equation*}
for ${\g k}={\g p}={\g q}$, that is responsible for splitting between Hubbard bands and eventually opening a Mott gap once the interactions are strong enough \cite{HK,Nature1,Nature2}. At the same time, we underline that the presence of the splitting between Hubbard bands by HK interaction for arbitrarily small amplitudes $U$ has no analogy in the system with Hubbard interaction where Fermi-liquid behavior is expected.

% $+\Big]$
% 
% 
% 
% 
% 
%  
% 
% 
% 
% , and   $U$ is amplitude of the many-body interactions. 
% 
% enriched by Hatsugai-Kohmoto interaction \cite{HK} $ 
% &+U\sum_kn_{{\g k}\uparrow}n_{{\g k}\downarrow}$
\section{Topology of many-body groundstate}
\subsection{Chern number of the groundstate}
Our Hamiltonian can be efficiently expressed in RBZ as 
\begin{equation}
   \mathcal{H}= \sum_{{\g k}\in{\rm RBZ}}\sum_{n\in\{0-4\}}|\hat \alpha^n_{\g k}\rangle  \mathcal{\hat H}^n_{\g k} \langle\hat\alpha^n_{\g k}|
\end{equation}
where $|\hat\alpha_{\g k}^n\rangle$ are vectors of Fock base-states in which Hamiltonian $\mathcal{H}$ can be exactly diagonalized, and where $n$ stands for the number of particles described by these states. Explicitly, we define  $|\hat\alpha_{\g k}^n\rangle$ as
% in the Fock space with distinguished not-mixing sectors with a given integer number of particles for a given $\g k$ in RBZ, 
 \begin{equation}
     \begin{split}
         |\hat\alpha_{{\g k}}^0\rangle =&\{\ket{0_{\g k};\!0_{\g k+Q}}\}\\
         |\hat\alpha_{{\g k}}^1\rangle =&\{\ket{\uparrow;\!0},\ket{\downarrow;\!0},\ket{0;\!\uparrow},\ket{0;\!\downarrow}\}\\
         |\hat\alpha_{{\g k}}^2\rangle =&\{\ket{\uparrow \downarrow;\!0}, \ket{0;\!\uparrow \downarrow},\ket{ \downarrow;\!  \downarrow},\ket{ \downarrow;\! \uparrow },\ket{ \uparrow;\! \uparrow },\ket{ \uparrow;\! \downarrow }\}\\
|\hat\alpha_{{\g k}}^3\rangle =&\{\ket{\uparrow \downarrow;\! \uparrow },\ket{\uparrow \downarrow;\! \downarrow},\ket{\uparrow ;\! \uparrow \downarrow},\ket{ \downarrow;\! \uparrow \downarrow}\}\\
|\hat\alpha_{{\g k}}^4\rangle =& \{\ket{\uparrow \downarrow;\! \uparrow \downarrow}\}.
 \end{split}
 \end{equation}
On the other hand Hamiltonian sectors $\mathcal{\hat H}^n_{\g k}$ in the above base apart from trivial $\mathcal{H}^0_{\g k}=\mathcal{H}^4_{\g k}=0$ are defined as
\begin{equation}
    \begin{split}
       \mathcal{\hat H}_{\g k}^1=&\sigma_z\otimes(\epsilon_{\g k}\mathbb{1}_2+\Re(g_{\g k})\sigma_x\!+\!\Im(g_{\g k})\sigma_y)+h\mathbb{1}_2\otimes\sigma_z\\
       &+V\sigma_x\otimes\mathbb{1}_2 -\frac{U}{2}\mathbb{1}_4 \\
  \mathcal{\hat H}_{\g k}^3 =&\sigma_z\otimes(\epsilon_{\g k}\mathbb{1}_2-\Re(g_{\g k})\sigma_x\!-\!\Im(g_{\g k})\sigma_y)+h\mathbb{1}_2\otimes\sigma_z\\
  &-V\sigma_x\otimes\mathbb{1}_2 -\frac{U}{2}\mathbb{1}_4  
    \end{split}
\end{equation}
and
  % \widetext
\begin{equation}
     \mathcal{\hat H}_{\g k}^2=\begin{pmatrix}
      2 \epsilon_{\g k}&0&0&-V&0&V\\
      0&-2 \epsilon_{\g k}&0&-V&0&V\\
      0&0&-2h-U &-g_{{\g k}} &0&g_{{\g k}}\\
      -V&-V&-g_{{\g k}}^*&-U   &g_{{\g k}}&0\\
      0&0&0&g_{{\g k}}^*&2h\!-\!U  &\!-g_{{\g k}}^*\\
      V&V& g_{{\g k}}^*&0 &\!-g_{{\g k}}&-U
    \end{pmatrix}.
\end{equation}
   % \endwidetext

\begin{figure*}[t]
    \centering   \includegraphics[width= \textwidth]{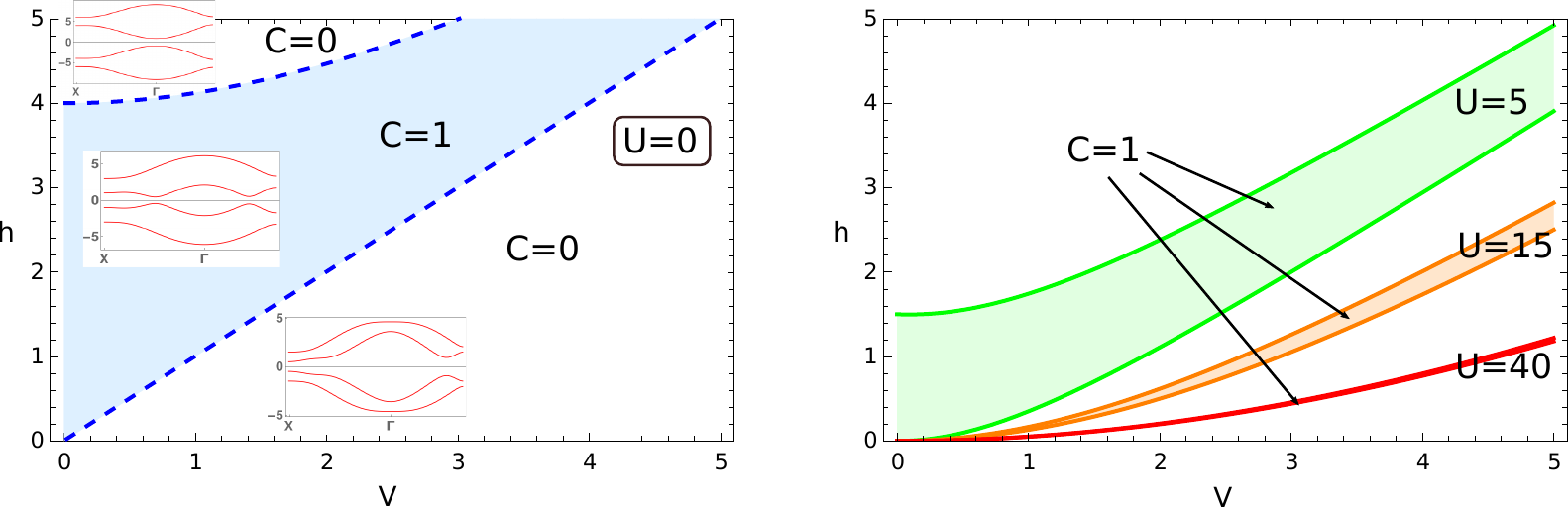} 
    \caption{Topological phase diagrams ("C" on the plots denotes Chern number) on $V-h$ plane for (left panel) uncorrelated system and (right panel) correlated systems characterized with selected values of interaction amplitude $U$. Phase boundaries marked with dashed lines denote standard topological phase transition, while those with solid lines denote  topological phase transitions at which spectral gap doesn't close (see Sec. IV). On the left panel we have included exemplary band structures for the three phases. }
    \label{num}
\end{figure*}

For the reason that the chemical potential, $\mu=U/2$  sets the filling in our system  to 2 not only on average but at each $\g k$ in RBZ, the global many-body groundstate is  the groundstate $|\psi_g({\g k})\rangle$ of $\mathcal{\hat H}^2_{\g k} $.
Therefore, here we calculate Chern number of the whole system as the one corresponding to $|\psi_g({\g k})\rangle$, i.e
\begin{equation}
\begin{split}
    \mathcal{C}=\frac{1}{2\pi i}\int_{{\g k}\in {\rm RBZ}}\! \!\!\!\!\!\!\!\!\!\!\!\!d^2 {\g k}\,\,\big( \partial_{k_x}\langle \psi_g&({\g k})|\partial_{k_y}|\psi_g({\g k})\rangle\\
    &-\partial_{k_y}\langle \psi_g({\g k})|\partial_{k_x}|\psi_g(k)\rangle\big)
    \end{split}
\end{equation}
The above topological invariant is directly linked to the anomalous Hall conductance, as shown in Ref. \cite{Thouless1985}.
In Fig.\ref{num} we show phase diagrams obtained  by direct integration with the use of discretization of RBZ \cite{Fukui_2005}. We note that although presence of the Rashba spin-orbit coupling mixing spin channels is critical for the onset of QAH phase, the topological phase boundaries do not change when the value of $\alpha$ is varied.  We rigorously support this statement in the next subsection. Nevertheless, for all numerical calculations we set  $\alpha=0.5$. 

Furthermore, as it can be seen in Fig.\ref{num}, even for extremely large values of interaction amplitude Chern insulator survives, though in a very restricted region that asymptotically reduces to a line. In the following we analyze that behavior on the analytical ground.

\subsection{Parity eigenvalues analysis}
Analysis of $ \mathcal{\hat H}_{\g k}^2$ at time reversal invariant momenta (TRIM) ${\g k}_{i}^*$, due to presence of inversion symmetry, can be useful in determination of borders between topological and trivial phases, provided that they differ by the parity of the Chern number. Here this seems sufficient as   our numerical analysis unveils the presence of only one type of topological phase   (cf. Fig. \ref{num}) with ${\cal C}=1$. 
 
First, we find eigenstates $|n_{\g k^*_i}\rangle$  and eigenvalues $E_{n\g k^*_i}$ of $ \mathcal{\hat H}^2_{{\g k}^*_{i}}$, where TRIM in RBZ can be chosen as ${\g k}^*_{i}\in\{\Gamma,X\}$.  Two lowest energy states  are  $|1_{\g k^*_i}\rangle\equiv|\downarrow,\downarrow\rangle$ with the eigenvalue $E_{1{\g k}^*_{i}}= -U-2h$ and $|2_{\g k^*_i}\rangle$ with eigenvalue $E_{2\g k^*_i}$;  the latter being the lowest eigenvalue of the matrix
    \begin{equation}
    \begin{split}&
   \begin{pmatrix}
        0 &\sqrt{2}\epsilon_{\g k^*_i}u_-&\sqrt{2}\epsilon_{\g k^*_i}u_+\\
       \sqrt{2}\epsilon_{\g k^*_i}u_-&\frac{-U-\sqrt{U^2+16V^2}}{2} &0\\
        \sqrt{2}\epsilon_{\g k^*_i}u_+&0&\frac{-U+\sqrt{U^2+16V^2}}{2}
    \end{pmatrix},\\
    &u_\pm= \sqrt{ 1\pm\frac{U}{\sqrt{U^2+16V^2}} },
    \label{ope}
    \end{split}
\end{equation} 
that is simply one of the blocks from block-diagonal form of $ \mathcal{\hat H}^2_{{\g k}^*_{i}}$ obtained with a suitable unitary transformation and spanned by  $\{|\uparrow\downarrow,0\rangle,|0,\uparrow\downarrow\rangle, \frac{1}{\sqrt{2}}(|\uparrow,\downarrow\rangle-|\downarrow,\uparrow\rangle)\}$ Fock states.  

At TRIM  states $|n_{\g k^*_i}\rangle$ are also eigenstates with eigenvalues $\eta_n$ of parity operator 
\begin{equation}
    P=\begin{pmatrix}
        1&0&0&0&0&0\\
        0&1&0&0&0&0\\
        0&0&-1&0&0&0\\
        0&0&0&1&0&0\\
        0&0&0&0&-1&0\\
        0&0&0&0&0&1
    \end{pmatrix}
\end{equation}
as $P\mathcal{\hat H}_{\g k}^2P=\mathcal{\hat H}_{-\g k}^2$ implies that $[P,\mathcal{\hat H}_{\g k^*}^2]=0$. At both TRIM we have that  $\eta_1=-1$ and $\eta_2=1$. Formally,  by defining  
\begin{equation}
   (-1)^\nu=\prod_{i\in{\Gamma,X}}{\rm sgn}[E_{1k^*_i}-E_{2k^*_i}] .\label{topo}
\end{equation}
$\nu=0$ signals trivial insulator and $\nu=1$ topologically nontrivial state. 
In other words, phase transition between trivial and nontrivial topological states takes place when groundstate and first excited state, having opposite parity, interchange. Note that, although presence of Rashba spin-orbit entails presence of states with opposite parity, its amplitude does not enter Eq. \eqref{topo}. This proves our statement from previous subsection that boundaries between topological and trivial phases do not depend on Rashba spin-orbit amplitude.

Leveraging supplementary topological invariant \eqref{topo} allowing for straightforward discrimination between topological and trivial phases in the following we  analyze specific phase boundaries displayed in Fig. \ref{num}. 

First, we examine conditions for transition from the trivial insulator to the Chern insulator with increasing Zeeman splitting $h$, i.e. from the "bottom" - cf. Fig. \ref{num}. At critical Zeeman splitting which we denote $h_b$ two lowest energy levels at TRIM $X$ cross. 
At this momenta these energy levels are found exactly:  $E_{1X}=-U-2h$ and $E_{2X}=-\frac{1}{2}(U+\sqrt{U^2+16V^2})$. Condition that $E_{1X}=E_{2X}$ sets the {\it bottom}  topological phase transition as
\begin{equation}
  h_b(V,U)=\frac{1}{4}(\sqrt{U^2+16 V^2}-U).
  \label{hb}
\end{equation}

  Second, topological phase transition into the same   Chern insulator phase but with decreasing $h$ (from the "top" - cf. Fig. \ref{num}) takes place at critical $h=h_t$ when $E_{1\Gamma}=E_{2\Gamma}$, i.e. is associated with crossing between lowest lying energy levels at $\Gamma$ TRIM.   Here $E_{2\Gamma}$ can be obtained exactly only for $U=0$ case, for which we find  $E_{2\Gamma}=-2\sqrt{V^2+\epsilon_\Gamma^2}$ (lowest eigenvalue of \eqref{ope}), where $\epsilon_{\Gamma}\equiv \epsilon_{{\g k}^*_\Gamma}$. Thus critical Zeeman splitting for the "top" topological phase transition in uncorrelated  case reads
 ${h_t(V,U=0)=\sqrt{V^2+\epsilon_\Gamma^2}}$. 
  On the other hand, in the opposite limit (i.e. large interaction limit, $\sqrt{2}\epsilon_\Gamma/U\ll1$), we resort to the second order perturbation theory when obtaining 
  $E_{2\Gamma}$ and we find 
  \begin{equation}
      E_{2\Gamma}\simeq E_{2X}-\frac{(\sqrt{U^2+16V^2}-U)(2\epsilon_\Gamma)^2}{\sqrt{U^2+16V^2}(\sqrt{U^2+16V^2}+U)}.
  \end{equation}
  Consequently,  approximate critical Zeeman splitting for the "top" topological phase transition in the strong interaction limit reads
  \begin{equation}
      \begin{split}
        h_t(V,U\!\gg\!\sqrt{2}\epsilon_\Gamma)&= h_b(V,U)\\
        &+\!\frac{(\sqrt{U^2+16V^2}-U)2\epsilon_\Gamma^2}{\sqrt{U^2\!+\!16V^2}(\sqrt{U^2\!+\!16V^2}\!+\!U)}
      \end{split}
  \end{equation}

In result, region on $V-h$ plane where Chern insulator with $\mathcal{C}=1$ is realized, lies between boundaries set by $h_b(U,V)$ and (if $U$ is sufficiently large) $h_t(U,V)$, and only asymptotically vanish when $U\to \infty$, while for large but finite $U$ forms a sharp edge (cf. Fig. \ref{num}). 

\section{Single particle Green's function properties}

Knowledge of the exact groundstate wave function of the considered  system is a necessary factor enabling for the analysis presented in the previous section. Nevertheless, in generic situations with more realistic many-body interactions the system properties are usually captured at the level of single particle Green's function. Therefore, topological invariant defined through single-particle Green's function \cite{Zhang_2012_2} is considered as the default one when it comes to description of the topological properties of the many-body system.
However, it has been recently proved \cite{Yang2019} that it needs to be treated with care as it can give false positive topological properties even if the system is trivial. For that reason, in the following we analyze topological and spectral properties of the considered system at the level of single-particle Green's function and carefully compare findings with the precise analysis of the many-body groundstate.

\subsection{Spectral function and absence of gap closing}
In the previous section we have shown that the topological phase transition  is signalled by crossing between ground state and first excited state, both many-body Fock states, that takes place at one of the time-reversal invariant momenta. As we are going to analyze in this section, this usual and expected behavior,  due to presence of correlations, does not entail closure of the spectral gap.

We first define
\begin{equation}
    \mathcal{\hat H}_{\g k}=\bigoplus_{n=0}^4 \mathcal{\hat H}^n_{\g k}\,\,\,\,\,\,\,\,\,\,\,\, {\rm and}\,\,\,\,\,\,\,\,\,\,\,\,  |\hat\alpha_{\g k}\rangle= \bigoplus_{n=0}^4 |\hat \alpha_{\g k}^n\rangle.
\end{equation}
Now, we can find unitary transformation $T_{\g k}$, such that  $T_{\g k}^\dagger\mathcal{\hat H}_{\g k}T_{\g k}=\hat E_{\g k}$ is diagonal. In result we can obtain spectral function as
\begin{equation}
    A({\g k},\omega)=\frac{1}{\pi}\Im\, {\rm Tr} [ \hat G_{\g k}(\omega-i\delta) ]
\end{equation}
where Matsubara Green's function  reads
\begin{equation}
    \begin{split}
       \hat G_{\g k}(i \omega ) =\begin{pmatrix}
     G^{1,1}_{\uparrow,\uparrow}&  G^{1,1}_{\uparrow,\downarrow}&  G^{1,2}_{\uparrow,\uparrow}&  G^{1,2}_{\uparrow,\downarrow}\\
      G^{1,1}_{\downarrow,\uparrow} & G^{1,1}_{\downarrow,\downarrow} &  G^{1,2}_{\downarrow,\uparrow} & G^{1,2}_{\downarrow,\downarrow}\\
   G^{2,1}_{\uparrow,\uparrow}& G^{2,1}_{\uparrow,\downarrow}&   G^{2,2}_{\uparrow,\uparrow}& G^{2,2}_{\uparrow,\downarrow}  \\
    G^{2,1}_{\downarrow,\uparrow} & G^{2,1}_{\downarrow,\downarrow} &   G^{2,2}_{\downarrow,\uparrow} & G^{2,2}_{\downarrow,\downarrow} \\
        \end{pmatrix}\label{gre}
    \end{split}    
\end{equation}
with matrix elements obtained from Lehmann representation as
\begin{equation}
    \begin{split}
         G^{n,n'}_{\sigma,\sigma'}(i\omega)= \sum_m\big[&\frac{\langle 0|c_{n,\sigma}|m\rangle\langle m|c^\dagger_{n',\sigma'}|0\rangle}{i\omega-(E_m-E_0)} \\ &+\frac{\langle m|c_{n,\sigma}|0\rangle\langle 0|c^\dagger_{n',\sigma'}|m\rangle}{i\omega+(E_m-E_0)}\big].
    \end{split}
\end{equation} 
In above $|m\rangle$ are the exact eigenvectors ($|0\rangle$ being groundstate)  of  $\mathcal{\hat H}_{\g k}$ with eigenvalues $E_m$ and 
 $c_{1\sigma}\equiv c_{{\g k}\sigma}$ and $c_{2\sigma}\equiv c_{{\g k+Q}\sigma}$.

\begin{figure}
    \centering
    \includegraphics[width=0.45\textwidth]{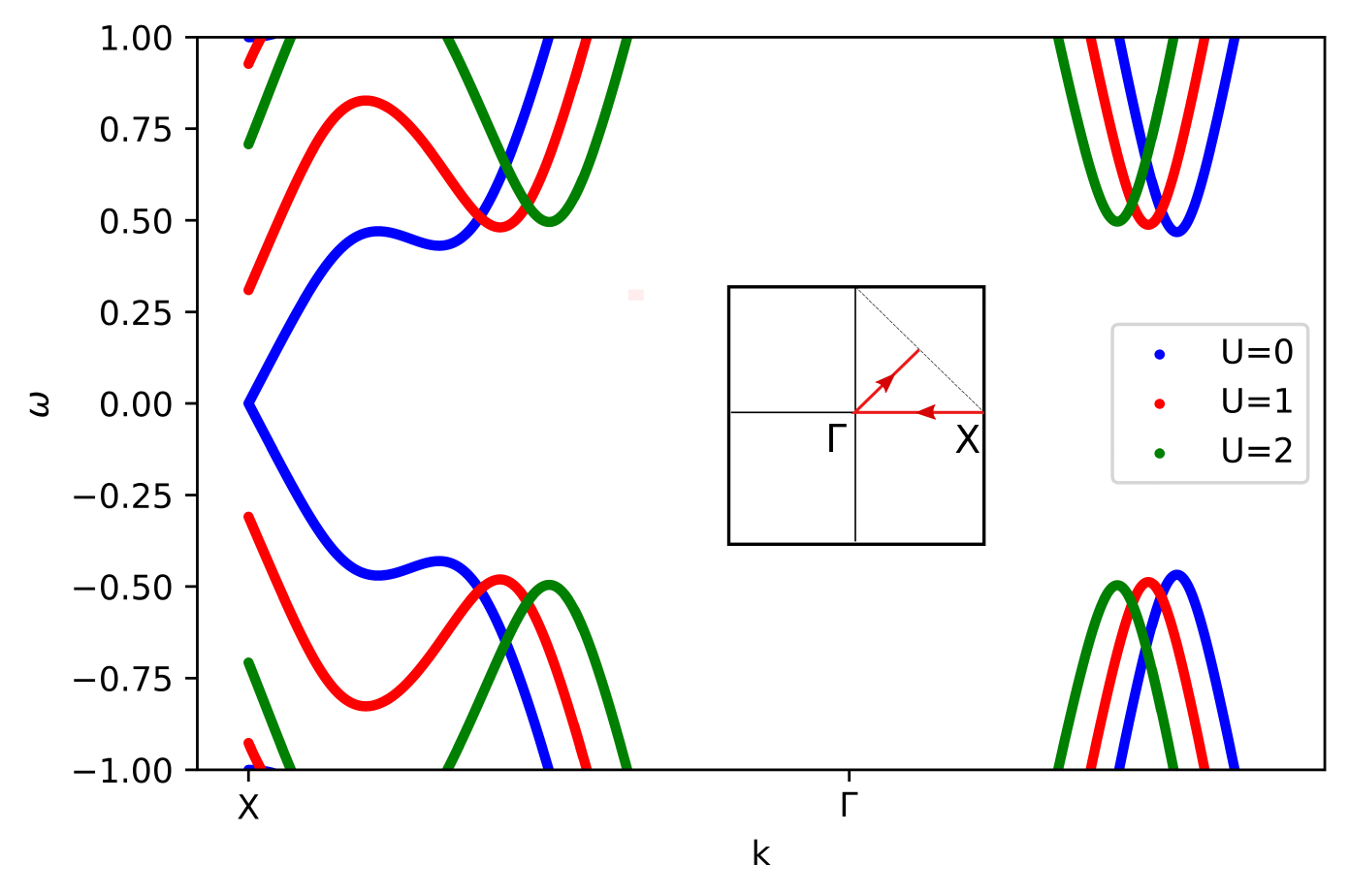}
    \caption{Plot of the spectral function in the vicinity of the Fermi level along a high symmetry path (cf. inset) for values of $h=h_b$, i.e. at the topological phase transition. Note the lack of a gap closing in the spectral function for non-zero interaction strength. The spectral weight for bands visible in the plot is close to 1.}
    \label{fig2}
\end{figure}

In the Fig. \ref{fig2} we show spectral function around the Fermi level for parameters providing  "bottom" topological phase transition, i.e. for $h=h_b(V,U)$, for three selected cases:  uncorrelated, $U=0$ and  correlated, $U=1$ and $U=2$ ones. It is evident that the spectral gap closes at the topological phase transition only for uncorrelated states. Such observation is consistent with a phenomenon called {\it first-order topological transition} found in other many-body topological systems with more realistic, usually Hubbard-type   interactions \cite{Budich2013,Amaricci_2015,Amaricci_2016,Troyer2016,Roy2016,Barbarino2019} for which  it has been also found that correlations can entail absence of the spectral gap closing on the topological transition. Here, relying on the previous analysis of the exact groundstate we shall simply explain microscopic reasons for such a situation and briefly elaborate on the misleading notion coined for this phenomenon.

Spectral function is a single-particle property of the system and as such tells us how the injected  electron behaves in our system. That means that the spectral function holds information about processes that involve excitation that change the number of particles in the system. On the other hand, as our analysis of the exact many-body groundstate unveils, the first excited state in the presence of interactions does not involve change of electron number. Therefore, spectral function is unable to  
capture crossing of the groundstate and first excited state that signals regular continuous topological phase transition without any feature suggesting first-order character.

In this light, given that spectral function already can provide misleading signatures on the character of topological transition,  there is a natural question whether single particle Green's function in a correlated state beyond Fermi liquid theory still contains enough information to calculate correct boundaries of topological phase. In the next subsection we affirmatively answer this question. 

\subsection{Topological invariant from Green's function}
Here we calculate the topological invariant of our system, following the approach in Ref. \onlinecite{Zhang_2012_2}. In general words, in this method one can treat the inverse of Green's function for zero frequency as an effective low-energy single-particle Hamiltonian,
\begin{equation}
H^{eff}_{\g k}=-G_{\g k}(0)^{-1}
\end{equation}
%\begin{equation}
  % H^{eff}=\sum_{{\g k}\in{\rm RBZ}} H^{eff}_{\g k}=-\sum_{{\g k}\in{\rm RBZ}}G_{\g k}(0)^{-1}
%\end{equation}
In such a situation determination of the topological invariant of the system reduces to analysis of $H^{eff}$ (which is Hermitian - cf. Eq. \eqref{gre}).

We first find eigenvectors and eigenvalues of effective Hamiltonian
\begin{equation}
    H^{eff}_{\g k}|n({\g k})\rangle=\mu_n({\g k})|n({\g k})\rangle.
\end{equation}
Then the many-body topological invariant can be obtained as
\begin{equation}
    \mathcal{\tilde C}=\frac{1}{2\pi}\int_{\rm RBZ} d^2k (\partial_{k_x}\mathcal{A}_y-\partial_{k_y}\mathcal{A}_x)
\end{equation}
and where $\mathcal{A}_i=-i\sum_{n(\mu_n<0)}\langle n({\g k})|\partial_{k_i}|n({\g k})\rangle$. 

We numerically confirm that the above approach provides the same topological phase borders as the approach utilizing true may-body groundstate presented in Sec III and in principle  $\mathcal{C}=\mathcal{\tilde C}$. We note that although here spectral function is unable to visualize inversion between groundstate and first excited state, single-particle Green's function holds correct information of the system's topology. 

\begin{figure}
    \centering
    \includegraphics[width=0.45\textwidth]{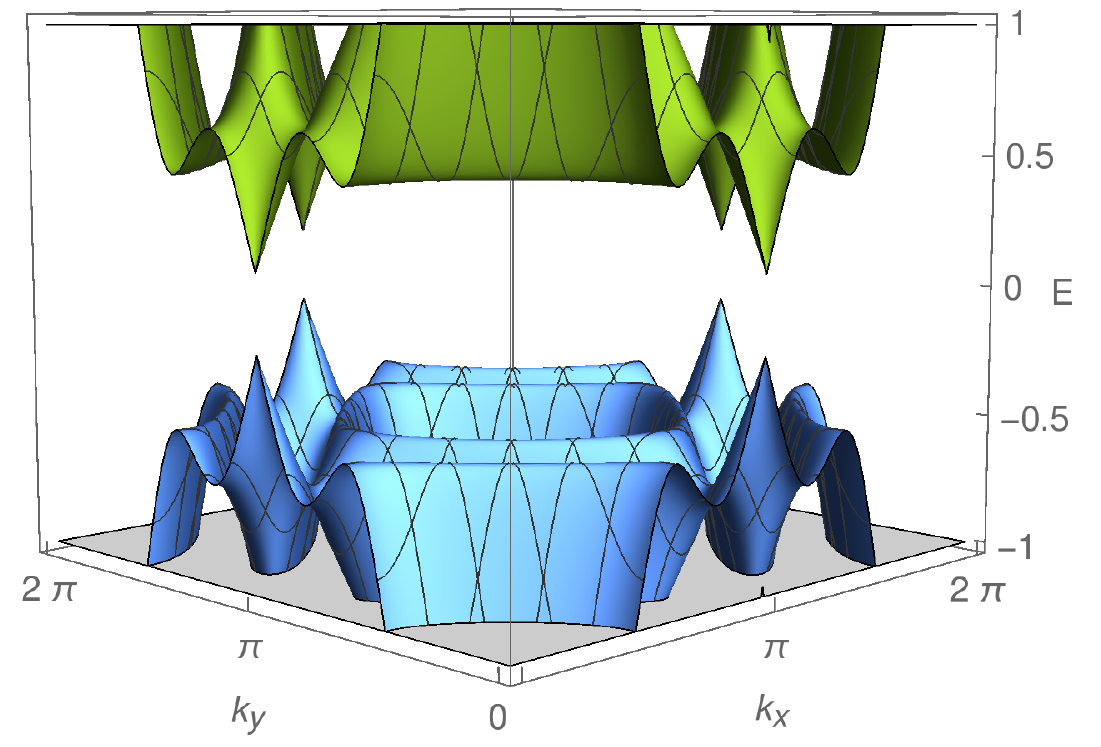}
    \caption{Eigenenergies of effective low-energy single particle Hamiltonian around Fermi level for parameters providing "bottom" topological phase transition, i.e.   $h=h_b$ (cf. Eq. \eqref{hb}). Note absence of the gap closing. }
    \label{fig3}
\end{figure}

\begin{figure}
    \centering
    \includegraphics[width=0.235\textwidth]{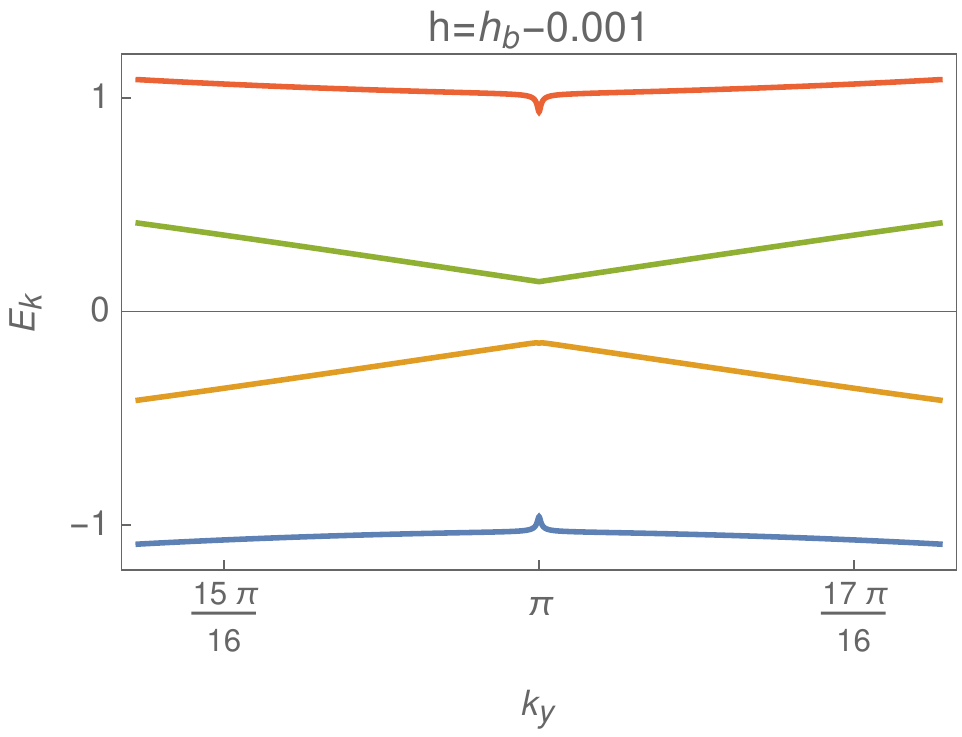}
    \includegraphics[width=0.235\textwidth]{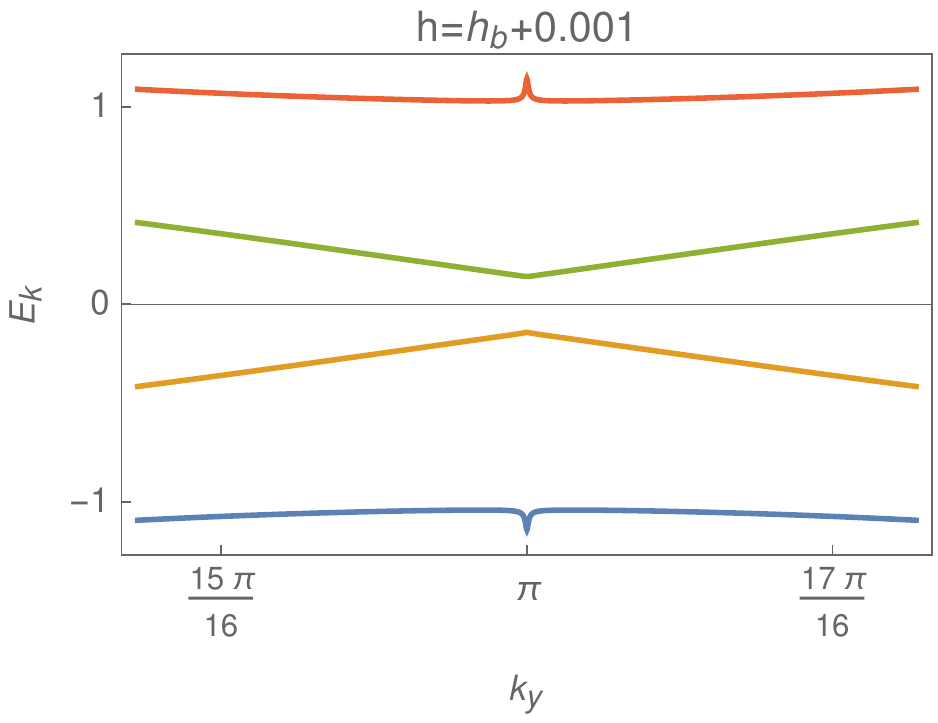}
    \caption{
    Bands of $H^{eff}_{\g k}$ around TRIM $X$ i.e. for $k_x=0$ (left panel) for  trivial phase ($h<h_b$) just before transition and (right panel) for topological phase ($h>h_b$) just after the transition. At the transition bands are characterized by a discontinuous jump at TRIM $X$.  }
    \label{fig5}
\end{figure}

In the following we shall analyze more deeply properties of the low-energy effective Hamiltonian. We found that the topological phase transition is again not associated with the bulk gap closing, as it is demonstrated in Fig. \ref{fig3}, and clearly the system develops singularity at $X$ TRIM.  To understand it in more details in Fig \ref{fig5} we plot  bands around TRIM $X$ just before and after transition what confirms expectation that the phase transition takes place through the discontinuous jump of eigen-energies at $X$. Particularly this feature has been responsible for calling such observation  {\it topological phase transition with first-order character} in the first place \cite{Amaricci_2015}.

\begin{figure}
    \centering
    \includegraphics[width=0.45\textwidth]{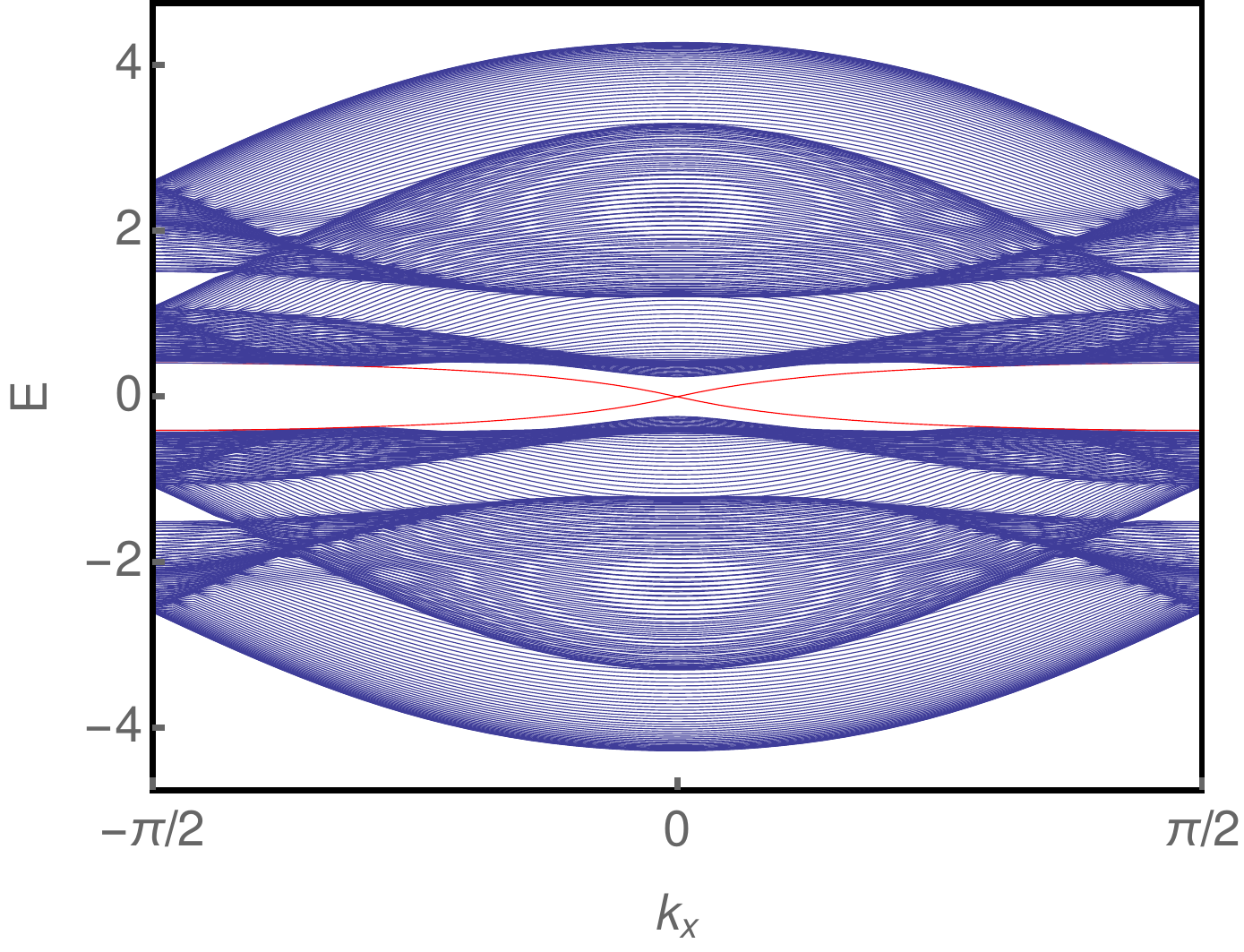}
    \caption{Bands of $H^{eff}_{\g k}$ with  open boundary conditions in $y$-direction in the topological phase. Edge states are clearly visible. The parametrs are $U=V=0.5$ and $h=h_b+0.1$.}
    \label{fig4}
\end{figure}

This brings us to a following question: if  topological and trivial phases are not separated by gap closing, is the topological phase inside the bulk separated from the trivial "outside"? To resolve this issue, in Fig. \ref{fig4} we plot bands of $H^{eff}_{\g k}$ but with the open boundaries along $y$-direction and we obtain a rather standard scenario. Namely, that edge states are not gaped and clearly connect bands from below the gap with these from above and their number per one edge agrees with the Chern number.

\subsection{Properties of $H^{eff}$ at TRIM}

In this section we shall again leverage inversion symmetry as a good tool for determining phase transition between trivial and topological phases \cite{Zhang_2012} with odd Chern number.
We start our ana\-lysis by finding the parity ope\-rator for the full Green's function,
\begin{equation}
    \mathcal{P}=\begin{pmatrix}
        1&0&0&0\\
        0&-1&0&0\\
        0&0&1&0\\
        0&0&0&-1
    \end{pmatrix}\label{parity}
\end{equation}
i.e. $\mathcal{P} \hat G_{\g k}(i \omega_m)\mathcal{P}= \hat G_{-\g k}(i \omega_m)$. 
The form of operator $\mathcal{P}$ follows from the symmetry of the Rashba coupling in $\mathcal{H}$.
Following approach presented in Refs. \onlinecite{Zhang_2012,Zhang_2012_2} we note that zero-frequency single-particle Green's function  at TRIMs is Hermitian, commutes with the parity operator and specifically for our situation reduces to
\begin{equation}
    \begin{split}
       \hat G_{{\g k}_i^*}(0) =\begin{pmatrix}
     G^{1,1}_{{\g k}_i^*\uparrow}(0)&  0&  G^{1,2}_{{\g k}_i^*\uparrow}(0)&  0\\
      0& G^{1,1}_{{\g k}_i^*\downarrow}(0) &  0 & G^{1,2}_{{\g k}_i^*\downarrow}(0)\\
   G^{2,1}_{{\g k}_i^*\uparrow}(0)& 0&   G^{2,2}_{{\g k}_i^*\uparrow}(0)& 0  \\
    0 & G^{2,1}_{{\g k}_i^*\downarrow}(0) &   0& G^{2,2}_{{\g k}_i^*\downarrow}(0) \\
        \end{pmatrix}
    \end{split}    
\end{equation}

\begin{figure}[t]
    \centering
    \includegraphics[width=0.45\textwidth]{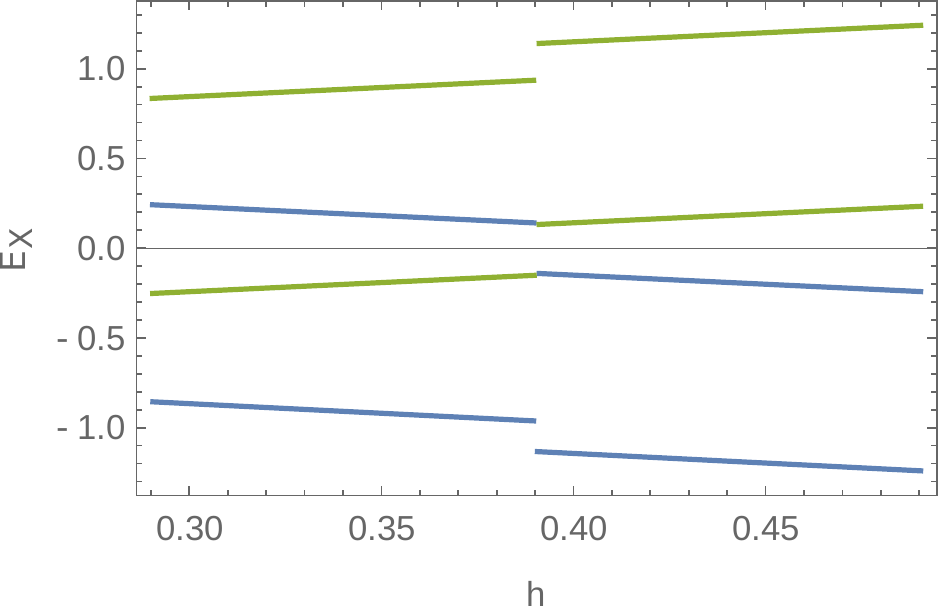}
    \caption{Evolution of eigen-energies of $H^{eff}_{\g k}$ at TRIM $X$ across the topological phase transition with increasing $h$. Color denotes different parities of the eigenstates. The topological phase transition takes place through the singularity instead of energy level crossing.  Parameters for the plot are $U=V=0.5$. }
    \label{fig6}
\end{figure}

Note that we can diagonalize above Green's function with a transformation that leaves the parity operator intact. Transformed and subsequently inverted Green's function defines effective Hamiltonian at TRIM
\begin{equation}
    H^{eff}_{{\g k}_i^*}= \begin{pmatrix}
        -\mu_{{\g k}_i^*\uparrow+}&0&0&0\\
        0&-\mu_{{\g k}_i^*\downarrow+}&0&0\\
        0&0&-\mu_{{\g k}_i^*\uparrow-}&0\\
        0&0&0&-\mu_{{\g k}_i^*\downarrow-}
    \end{pmatrix}
\end{equation}
where
\begin{equation}
    (\mu_{{\g k}_i^*\sigma\pm})^{-1}= \frac{G^{1,1}_{\sigma\g k^*}\!+\!G^{2,2}_{\sigma\g k^*}}{2}\pm\sqrt{\big(\frac{G^{1,1}_{\sigma\g k^*}\!-\!G^{2,2}_{\sigma\g k^*} }{2}\big)^2+|G^{1,2}_{\sigma\g k^*}|^2},
    \end{equation}
that still commutes with a parity operator Eq. \eqref{parity}, and therefore  both operators have common eigenvectors, i.e.,
\begin{equation}
\begin{split}
    H^{eff}_{\g k^*}|\sigma,\pm,\g k^*\rangle&=-\mu_{\sigma,\pm,\g k^*}|\sigma,\pm,\g k^*\rangle\\
 \mathcal{P}|\sigma,\pm,\g k^*\rangle&=\eta_{\sigma,\g k^*}  |\sigma,\pm,\g k^*\rangle
\end{split}
    \end{equation}
Because of the inversion symmetry \cite{Zhang_2012} we can define invariant $\nu$ by 
\begin{equation}
    (-1)^\nu\equiv\prod_{\g k^*}\prod_{\{\sigma,\pm\}\in {\rm R-zero}} \eta_{\sigma,\g k^*}^{1/2} 
\end{equation}
where ${\rm R-zero}$ is defined by eigenstates $|\sigma,\pm\rangle$ with negative eigenvalue $-\mu_{\sigma,\pm}<0$. As a result we find that the topological phase transition is signalled when invariant $\nu$ changes between 0 and 1, i.e., when  eigenvalues for states with different parities change sign at TRIM.
In Fig. \ref{fig6} we plot eigen-energies $\mu_{\sigma,\pm}$ for increasing field $h$ across the topological phase transition at $h=h_b$ and we found, peculiar for the single particle picture, discontinuous jump of eigen-energies at the topological phase transition, associated with a change of parities of states with positive energies. Such behavior can be understood on the following grounds. Although, at the topological phase transition, the ground-state energy remains continuous the state itself changes vastly due to the level crossing with the first excited state within the $|\hat\alpha_{{\g k}}^2\rangle$ manifold. This makes the spectrum of the single particle excitations discontinuous  as it is visible in Fig. \ref{fig6}.  

\section{Summary and conclusions}
In the present work we have proposed the model in which the interplay between ionic potential, Zeeman splitting and Rashba spin-orbit interaction realizes quantum anomalous Hall state. The model, up to the best of our knowledge has not been reported before to host topologically non-trivial state. In our opinion model has an advantage of being potentially realizable experimentally. In future works we are going to explore its potential to govern underlying physics of thin layers of semiconductors having rocksalt structure and negligible internal spin-orbit coupling, epitaxially grown on the substrate ensuring presence of Rashba interaction due to lack of spatial inversion,  and doped with magnetic ions providing sizable Zeeman splitting.

Moreover, we have studied the influence of the many-body interaction, introduced in the Hatsugai-Kohmoto spirit \cite{HK}, on the stability of the topological phase by two complementary techniques leveraging the closed form of the ground state and Green's function approach.  Interestingly, in the presence of correlations we have found that topological phase transition accompanied with expected crossing between groundstate and first excited state at the same time is characterized with the absence of a gap closing in the spectral function.
The latter feature  has been, in light of our results, misleadingly called {\it first order topological transition}. 
We explain the absence of spectral gap closing at the topological transition  by referring to the many-body nature of  lowest energy excited states. To analyze this feature further, by considering the inverse of zero frequency single-particle Green's function as an effective low-energy Hamiltonian, we demonstrated that topological phase transition takes place through a band discontinuity at the time reversal invariant momenta. Thus our work constitutes another example \cite{Yang2019} where analysis of single-particle Green's function in a correlated system can provide misleading conclusions related to topological properties.

\section*{Acknowledgments}  
The work is supported by the Foundation for Polish Science through the International Research
Agendas program co-financed by the European Union within the Smart Growth Operational Programme. W.B. acknowledges support by Narodowe Centrum Nauki 
(NCN, National Science Centre, Poland) Project No. 2019/34/E/ST3/00404.

%  \bibliography{rashba.bib}

\begin{thebibliography}{45}%
\makeatletter
\providecommand \@ifxundefined [1]{%
 \@ifx{#1\undefined}
}%
\providecommand \@ifnum [1]{%
 \ifnum #1\expandafter \@firstoftwo
 \else \expandafter \@secondoftwo
 \fi
}%
\providecommand \@ifx [1]{%
 \ifx #1\expandafter \@firstoftwo
 \else \expandafter \@secondoftwo
 \fi
}%
\providecommand \natexlab [1]{#1}%
\providecommand \enquote  [1]{``#1''}%
\providecommand \bibnamefont  [1]{#1}%
\providecommand \bibfnamefont [1]{#1}%
\providecommand \citenamefont [1]{#1}%
\providecommand \href@noop [0]{\@secondoftwo}%
\providecommand \href [0]{\begingroup \@sanitize@url \@href}%
\providecommand \@href[1]{\@@startlink{#1}\@@href}%
\providecommand \@@href[1]{\endgroup#1\@@endlink}%
\providecommand \@sanitize@url [0]{\catcode `\\12\catcode `\$12\catcode
  `\&12\catcode `\#12\catcode `\^12\catcode `\_12\catcode `\%12\relax}%
\providecommand \@@startlink[1]{}%
\providecommand \@@endlink[0]{}%
\providecommand \url  [0]{\begingroup\@sanitize@url \@url }%
\providecommand \@url [1]{\endgroup\@href {#1}{\urlprefix }}%
\providecommand \urlprefix  [0]{URL }%
\providecommand \Eprint [0]{\href }%
\providecommand \doibase [0]{https://doi.org/}%
\providecommand \selectlanguage [0]{\@gobble}%
\providecommand \bibinfo  [0]{\@secondoftwo}%
\providecommand \bibfield  [0]{\@secondoftwo}%
\providecommand \translation [1]{[#1]}%
\providecommand \BibitemOpen [0]{}%
\providecommand \bibitemStop [0]{}%
\providecommand \bibitemNoStop [0]{.\EOS\space}%
\providecommand \EOS [0]{\spacefactor3000\relax}%
\providecommand \BibitemShut  [1]{\csname bibitem#1\endcsname}%
\let\auto@bib@innerbib\@empty
%</preamble>
\bibitem [{\citenamefont {Rachel}(2018)}]{Rachel_2018}%
  \BibitemOpen
  \bibfield  {author} {\bibinfo {author} {\bibfnamefont {S.}~\bibnamefont
  {Rachel}},\ }\bibfield  {title} {\bibinfo {title} {Interacting topological
  insulators: a review},\ }\href {https://doi.org/10.1088/1361-6633/aad6a6}
  {\bibfield  {journal} {\bibinfo  {journal} {Reports on Progress in Physics}\
  }\textbf {\bibinfo {volume} {81}},\ \bibinfo {pages} {116501} (\bibinfo
  {year} {2018})}\BibitemShut {NoStop}%
\bibitem [{\citenamefont {Stoudenmire}\ \emph {et~al.}(2011)\citenamefont
  {Stoudenmire}, \citenamefont {Alicea}, \citenamefont {Starykh},\ and\
  \citenamefont {Fisher}}]{Fisher2011}%
  \BibitemOpen
  \bibfield  {author} {\bibinfo {author} {\bibfnamefont {E.~M.}\ \bibnamefont
  {Stoudenmire}}, \bibinfo {author} {\bibfnamefont {J.}~\bibnamefont {Alicea}},
  \bibinfo {author} {\bibfnamefont {O.~A.}\ \bibnamefont {Starykh}},\ and\
  \bibinfo {author} {\bibfnamefont {M.~P.}\ \bibnamefont {Fisher}},\ }\bibfield
   {title} {\bibinfo {title} {Interaction effects in topological
  superconducting wires supporting \mbox{Majorana} fermions},\ }\href
  {https://doi.org/10.1103/PhysRevB.84.014503} {\bibfield  {journal} {\bibinfo
  {journal} {Phys. Rev. B}\ }\textbf {\bibinfo {volume} {84}},\ \bibinfo
  {pages} {014503} (\bibinfo {year} {2011})}\BibitemShut {NoStop}%
\bibitem [{\citenamefont {Hohenadler}\ and\ \citenamefont
  {Assaad}(2013)}]{Assaad2013}%
  \BibitemOpen
  \bibfield  {author} {\bibinfo {author} {\bibfnamefont {M.}~\bibnamefont
  {Hohenadler}}\ and\ \bibinfo {author} {\bibfnamefont {F.~F.}\ \bibnamefont
  {Assaad}},\ }\bibfield  {title} {\bibinfo {title} {Correlation effects in
  two-dimensional topological insulators},\ }\href
  {https://doi.org/10.1088/0953-8984/25/14/143201} {\bibfield  {journal}
  {\bibinfo  {journal} {Journal of Physics: Condensed Matter}\ }\textbf
  {\bibinfo {volume} {25}},\ \bibinfo {pages} {143201} (\bibinfo {year}
  {2013})}\BibitemShut {NoStop}%
\bibitem [{\citenamefont {Wieckowski}\ \emph {et~al.}(2018)\citenamefont
  {Wieckowski}, \citenamefont {Ma\ifmmode~\acute{s}\else \'{s}\fi{}ka},\ and\
  \citenamefont {Mierzejewski}}]{Maska2018}%
  \BibitemOpen
  \bibfield  {author} {\bibinfo {author} {\bibfnamefont {A.}~\bibnamefont
  {Wieckowski}}, \bibinfo {author} {\bibfnamefont {M.~M.}\ \bibnamefont
  {Ma\ifmmode~\acute{s}\else \'{s}\fi{}ka}},\ and\ \bibinfo {author}
  {\bibfnamefont {M.}~\bibnamefont {Mierzejewski}},\ }\bibfield  {title}
  {\bibinfo {title} {Identification of \mbox{Majorana} modes in interacting
  systems by local integrals of motion},\ }\href
  {https://doi.org/10.1103/PhysRevLett.120.040504} {\bibfield  {journal}
  {\bibinfo  {journal} {Phys. Rev. Lett.}\ }\textbf {\bibinfo {volume} {120}},\
  \bibinfo {pages} {040504} (\bibinfo {year} {2018})}\BibitemShut {NoStop}%
\bibitem [{\citenamefont {Altland}\ and\ \citenamefont
  {Zirnbauer}(1997)}]{atland1997}%
  \BibitemOpen
  \bibfield  {author} {\bibinfo {author} {\bibfnamefont {A.}~\bibnamefont
  {Altland}}\ and\ \bibinfo {author} {\bibfnamefont {M.~R.}\ \bibnamefont
  {Zirnbauer}},\ }\bibfield  {title} {\bibinfo {title} {Nonstandard symmetry
  classes in mesoscopic normal-superconducting hybrid structures},\ }\href
  {https://doi.org/10.1103/PhysRevB.55.1142} {\bibfield  {journal} {\bibinfo
  {journal} {Phys. Rev. B}\ }\textbf {\bibinfo {volume} {55}},\ \bibinfo
  {pages} {1142} (\bibinfo {year} {1997})}\BibitemShut {NoStop}%
\bibitem [{\citenamefont {Chiu}\ \emph {et~al.}(2013)\citenamefont {Chiu},
  \citenamefont {Yao},\ and\ \citenamefont {Ryu}}]{Chiu13}%
  \BibitemOpen
  \bibfield  {author} {\bibinfo {author} {\bibfnamefont {C.-K.}\ \bibnamefont
  {Chiu}}, \bibinfo {author} {\bibfnamefont {H.}~\bibnamefont {Yao}},\ and\
  \bibinfo {author} {\bibfnamefont {S.}~\bibnamefont {Ryu}},\ }\bibfield
  {title} {\bibinfo {title} {Classification of topological insulators and
  superconductors in the presence of reflection symmetry},\ }\href
  {https://doi.org/10.1103/PhysRevB.88.075142} {\bibfield  {journal} {\bibinfo
  {journal} {Phys. Rev. B}\ }\textbf {\bibinfo {volume} {88}},\ \bibinfo
  {pages} {075142} (\bibinfo {year} {2013})}\BibitemShut {NoStop}%
\bibitem [{\citenamefont {Chiu}\ and\ \citenamefont {Schnyder}(2014)}]{Chiu14}%
  \BibitemOpen
  \bibfield  {author} {\bibinfo {author} {\bibfnamefont {C.-K.}\ \bibnamefont
  {Chiu}}\ and\ \bibinfo {author} {\bibfnamefont {A.~P.}\ \bibnamefont
  {Schnyder}},\ }\bibfield  {title} {\bibinfo {title} {Classification of
  reflection-symmetry-protected topological semimetals and nodal
  superconductors},\ }\href {https://doi.org/10.1103/PhysRevB.90.205136}
  {\bibfield  {journal} {\bibinfo  {journal} {Phys. Rev. B}\ }\textbf {\bibinfo
  {volume} {90}},\ \bibinfo {pages} {205136} (\bibinfo {year}
  {2014})}\BibitemShut {NoStop}%
\bibitem [{\citenamefont {Shiozaki}\ and\ \citenamefont {Sato}(2014)}]{Sato14}%
  \BibitemOpen
  \bibfield  {author} {\bibinfo {author} {\bibfnamefont {K.}~\bibnamefont
  {Shiozaki}}\ and\ \bibinfo {author} {\bibfnamefont {M.}~\bibnamefont
  {Sato}},\ }\bibfield  {title} {\bibinfo {title} {Topology of crystalline
  insulators and superconductors},\ }\href
  {https://doi.org/10.1103/PhysRevB.90.165114} {\bibfield  {journal} {\bibinfo
  {journal} {Phys. Rev. B}\ }\textbf {\bibinfo {volume} {90}},\ \bibinfo
  {pages} {165114} (\bibinfo {year} {2014})}\BibitemShut {NoStop}%
\bibitem [{\citenamefont {Shiozaki}\ \emph {et~al.}(2016)\citenamefont
  {Shiozaki}, \citenamefont {Sato},\ and\ \citenamefont {Gomi}}]{Shio16}%
  \BibitemOpen
  \bibfield  {author} {\bibinfo {author} {\bibfnamefont {K.}~\bibnamefont
  {Shiozaki}}, \bibinfo {author} {\bibfnamefont {M.}~\bibnamefont {Sato}},\
  and\ \bibinfo {author} {\bibfnamefont {K.}~\bibnamefont {Gomi}},\ }\bibfield
  {title} {\bibinfo {title} {Topology of nonsymmorphic crystalline insulators
  and superconductors},\ }\href {https://doi.org/10.1103/PhysRevB.93.195413}
  {\bibfield  {journal} {\bibinfo  {journal} {Phys. Rev. B}\ }\textbf {\bibinfo
  {volume} {93}},\ \bibinfo {pages} {195413} (\bibinfo {year}
  {2016})}\BibitemShut {NoStop}%
\bibitem [{\citenamefont {Budich}\ \emph {et~al.}(2013)\citenamefont {Budich},
  \citenamefont {Trauzettel},\ and\ \citenamefont {Sangiovanni}}]{Budich2013}%
  \BibitemOpen
  \bibfield  {author} {\bibinfo {author} {\bibfnamefont {J.~C.}\ \bibnamefont
  {Budich}}, \bibinfo {author} {\bibfnamefont {B.}~\bibnamefont {Trauzettel}},\
  and\ \bibinfo {author} {\bibfnamefont {G.}~\bibnamefont {Sangiovanni}},\
  }\bibfield  {title} {\bibinfo {title} {Fluctuation-driven topological
  \mbox{Hund} insulators},\ }\href {https://doi.org/10.1103/PhysRevB.87.235104}
  {\bibfield  {journal} {\bibinfo  {journal} {Phys. Rev. B}\ }\textbf {\bibinfo
  {volume} {87}},\ \bibinfo {pages} {235104} (\bibinfo {year}
  {2013})}\BibitemShut {NoStop}%
\bibitem [{\citenamefont {Amaricci}\ \emph {et~al.}(2015)\citenamefont
  {Amaricci}, \citenamefont {Budich}, \citenamefont {Capone}, \citenamefont
  {Trauzettel},\ and\ \citenamefont {Sangiovanni}}]{Amaricci_2015}%
  \BibitemOpen
  \bibfield  {author} {\bibinfo {author} {\bibfnamefont {A.}~\bibnamefont
  {Amaricci}}, \bibinfo {author} {\bibfnamefont {J.~C.}\ \bibnamefont
  {Budich}}, \bibinfo {author} {\bibfnamefont {M.}~\bibnamefont {Capone}},
  \bibinfo {author} {\bibfnamefont {B.}~\bibnamefont {Trauzettel}},\ and\
  \bibinfo {author} {\bibfnamefont {G.}~\bibnamefont {Sangiovanni}},\
  }\bibfield  {title} {\bibinfo {title} {First-order character and observable
  signatures of topological quantum phase transitions},\ }\href
  {https://doi.org/10.1103/PhysRevLett.114.185701} {\bibfield  {journal}
  {\bibinfo  {journal} {Phys. Rev. Lett.}\ }\textbf {\bibinfo {volume} {114}},\
  \bibinfo {pages} {185701} (\bibinfo {year} {2015})}\BibitemShut {NoStop}%
\bibitem [{\citenamefont {Amaricci}\ \emph {et~al.}(2016)\citenamefont
  {Amaricci}, \citenamefont {Budich}, \citenamefont {Capone}, \citenamefont
  {Trauzettel},\ and\ \citenamefont {Sangiovanni}}]{Amaricci_2016}%
  \BibitemOpen
  \bibfield  {author} {\bibinfo {author} {\bibfnamefont {A.}~\bibnamefont
  {Amaricci}}, \bibinfo {author} {\bibfnamefont {J.~C.}\ \bibnamefont
  {Budich}}, \bibinfo {author} {\bibfnamefont {M.}~\bibnamefont {Capone}},
  \bibinfo {author} {\bibfnamefont {B.}~\bibnamefont {Trauzettel}},\ and\
  \bibinfo {author} {\bibfnamefont {G.}~\bibnamefont {Sangiovanni}},\
  }\bibfield  {title} {\bibinfo {title} {Strong correlation effects on
  topological quantum phase transitions in three dimensions},\ }\href
  {https://doi.org/10.1103/PhysRevB.93.235112} {\bibfield  {journal} {\bibinfo
  {journal} {Phys. Rev. B}\ }\textbf {\bibinfo {volume} {93}},\ \bibinfo
  {pages} {235112} (\bibinfo {year} {2016})}\BibitemShut {NoStop}%
\bibitem [{\citenamefont {Imri\ifmmode~\check{s}\else \v{s}\fi{}ka}\ \emph
  {et~al.}(2016)\citenamefont {Imri\ifmmode~\check{s}\else \v{s}\fi{}ka},
  \citenamefont {Wang},\ and\ \citenamefont {Troyer}}]{Troyer2016}%
  \BibitemOpen
  \bibfield  {author} {\bibinfo {author} {\bibfnamefont {J.}~\bibnamefont
  {Imri\ifmmode~\check{s}\else \v{s}\fi{}ka}}, \bibinfo {author} {\bibfnamefont
  {L.}~\bibnamefont {Wang}},\ and\ \bibinfo {author} {\bibfnamefont
  {M.}~\bibnamefont {Troyer}},\ }\bibfield  {title} {\bibinfo {title}
  {First-order topological phase transition of the \mbox{Haldane-Hubbard}
  model},\ }\href {https://doi.org/10.1103/PhysRevB.94.035109} {\bibfield
  {journal} {\bibinfo  {journal} {Phys. Rev. B}\ }\textbf {\bibinfo {volume}
  {94}},\ \bibinfo {pages} {035109} (\bibinfo {year} {2016})}\BibitemShut
  {NoStop}%
\bibitem [{\citenamefont {Roy}\ \emph {et~al.}(2016)\citenamefont {Roy},
  \citenamefont {Goswami},\ and\ \citenamefont {Sau}}]{Roy2016}%
  \BibitemOpen
  \bibfield  {author} {\bibinfo {author} {\bibfnamefont {B.}~\bibnamefont
  {Roy}}, \bibinfo {author} {\bibfnamefont {P.}~\bibnamefont {Goswami}},\ and\
  \bibinfo {author} {\bibfnamefont {J.~D.}\ \bibnamefont {Sau}},\ }\bibfield
  {title} {\bibinfo {title} {Continuous and discontinuous topological quantum
  phase transitions},\ }\href {https://doi.org/10.1103/PhysRevB.94.041101}
  {\bibfield  {journal} {\bibinfo  {journal} {Phys. Rev. B}\ }\textbf {\bibinfo
  {volume} {94}},\ \bibinfo {pages} {041101} (\bibinfo {year}
  {2016})}\BibitemShut {NoStop}%
\bibitem [{\citenamefont {Barbarino}\ \emph {et~al.}(2019)\citenamefont
  {Barbarino}, \citenamefont {Sangiovanni},\ and\ \citenamefont
  {Budich}}]{Barbarino2019}%
  \BibitemOpen
  \bibfield  {author} {\bibinfo {author} {\bibfnamefont {S.}~\bibnamefont
  {Barbarino}}, \bibinfo {author} {\bibfnamefont {G.}~\bibnamefont
  {Sangiovanni}},\ and\ \bibinfo {author} {\bibfnamefont {J.~C.}\ \bibnamefont
  {Budich}},\ }\bibfield  {title} {\bibinfo {title} {First-order topological
  quantum phase transition in a strongly correlated ladder},\ }\href
  {https://doi.org/10.1103/PhysRevB.99.075158} {\bibfield  {journal} {\bibinfo
  {journal} {Phys. Rev. B}\ }\textbf {\bibinfo {volume} {99}},\ \bibinfo
  {pages} {075158} (\bibinfo {year} {2019})}\BibitemShut {NoStop}%
\bibitem [{\citenamefont {Dzero}\ \emph {et~al.}(2010)\citenamefont {Dzero},
  \citenamefont {Sun}, \citenamefont {Galitski},\ and\ \citenamefont
  {Coleman}}]{Coleman_2010}%
  \BibitemOpen
  \bibfield  {author} {\bibinfo {author} {\bibfnamefont {M.}~\bibnamefont
  {Dzero}}, \bibinfo {author} {\bibfnamefont {K.}~\bibnamefont {Sun}}, \bibinfo
  {author} {\bibfnamefont {V.}~\bibnamefont {Galitski}},\ and\ \bibinfo
  {author} {\bibfnamefont {P.}~\bibnamefont {Coleman}},\ }\bibfield  {title}
  {\bibinfo {title} {Topological \mbox{Kondo} insulators},\ }\href
  {https://doi.org/10.1103/PhysRevLett.104.106408} {\bibfield  {journal}
  {\bibinfo  {journal} {Phys. Rev. Lett.}\ }\textbf {\bibinfo {volume} {104}},\
  \bibinfo {pages} {106408} (\bibinfo {year} {2010})}\BibitemShut {NoStop}%
\bibitem [{\citenamefont {Wysoki\ifmmode~\acute{n}\else \'{n}\fi{}ski}\ and\
  \citenamefont {Fabrizio}(2016)}]{Wysokinski_2016}%
  \BibitemOpen
  \bibfield  {author} {\bibinfo {author} {\bibfnamefont {M.~M.}\ \bibnamefont
  {Wysoki\ifmmode~\acute{n}\else \'{n}\fi{}ski}}\ and\ \bibinfo {author}
  {\bibfnamefont {M.}~\bibnamefont {Fabrizio}},\ }\bibfield  {title} {\bibinfo
  {title} {Many-body breakdown of indirect gap in topological \mbox{Kondo}
  insulators},\ }\href {https://doi.org/10.1103/PhysRevB.94.121102} {\bibfield
  {journal} {\bibinfo  {journal} {Phys. Rev. B}\ }\textbf {\bibinfo {volume}
  {94}},\ \bibinfo {pages} {121102} (\bibinfo {year} {2016})}\BibitemShut
  {NoStop}%
\bibitem [{\citenamefont {Raghu}\ \emph {et~al.}(2008)\citenamefont {Raghu},
  \citenamefont {Qi}, \citenamefont {Honerkamp},\ and\ \citenamefont
  {Zhang}}]{Raghu_2008}%
  \BibitemOpen
  \bibfield  {author} {\bibinfo {author} {\bibfnamefont {S.}~\bibnamefont
  {Raghu}}, \bibinfo {author} {\bibfnamefont {X.-L.}\ \bibnamefont {Qi}},
  \bibinfo {author} {\bibfnamefont {C.}~\bibnamefont {Honerkamp}},\ and\
  \bibinfo {author} {\bibfnamefont {S.-C.}\ \bibnamefont {Zhang}},\ }\bibfield
  {title} {\bibinfo {title} {Topological \mbox{Mott} insulators},\ }\href
  {https://doi.org/10.1103/PhysRevLett.100.156401} {\bibfield  {journal}
  {\bibinfo  {journal} {Phys. Rev. Lett.}\ }\textbf {\bibinfo {volume} {100}},\
  \bibinfo {pages} {156401} (\bibinfo {year} {2008})}\BibitemShut {NoStop}%
\bibitem [{\citenamefont {Morimoto}\ and\ \citenamefont
  {Nagaosa}(2016)}]{Morimoto2016}%
  \BibitemOpen
  \bibfield  {author} {\bibinfo {author} {\bibfnamefont {T.}~\bibnamefont
  {Morimoto}}\ and\ \bibinfo {author} {\bibfnamefont {N.}~\bibnamefont
  {Nagaosa}},\ }\bibfield  {title} {\bibinfo {title} {Weyl \mbox{Mott}
  insulator},\ }\href {https://doi.org/10.1038/srep19853} {\bibfield  {journal}
  {\bibinfo  {journal} {Scientific Reports}\ }\textbf {\bibinfo {volume} {6}},\
  \bibinfo {pages} {19853} (\bibinfo {year} {2016})}\BibitemShut {NoStop}%
\bibitem [{\citenamefont {Yang}(2019)}]{Yang2019}%
  \BibitemOpen
  \bibfield  {author} {\bibinfo {author} {\bibfnamefont {M.-F.}\ \bibnamefont
  {Yang}},\ }\bibfield  {title} {\bibinfo {title} {Manifestation of topological
  behaviors in interacting weyl systems: One-body versus two-body
  correlations},\ }\href {https://doi.org/10.1103/PhysRevB.100.245137}
  {\bibfield  {journal} {\bibinfo  {journal} {Phys. Rev. B}\ }\textbf {\bibinfo
  {volume} {100}},\ \bibinfo {pages} {245137} (\bibinfo {year}
  {2019})}\BibitemShut {NoStop}%
\bibitem [{\citenamefont {Mai}\ \emph {et~al.}(2022{\natexlab{a}})\citenamefont
  {Mai}, \citenamefont {Feldman},\ and\ \citenamefont {Phillips}}]{arxiv1}%
  \BibitemOpen
  \bibfield  {author} {\bibinfo {author} {\bibfnamefont {P.}~\bibnamefont
  {Mai}}, \bibinfo {author} {\bibfnamefont {B.~E.}\ \bibnamefont {Feldman}},\
  and\ \bibinfo {author} {\bibfnamefont {P.~W.}\ \bibnamefont {Phillips}},\
  }\bibfield  {title} {\bibinfo {title} {Topological \mbox{Mott} insulator at
  quarter filling in the interacting \mbox{Haldane} model},\ }\href@noop {} {\
  (\bibinfo {year} {2022}{\natexlab{a}})},\ \Eprint
  {https://arxiv.org/abs/arXiv:2207.01638} {arXiv:2207.01638} \BibitemShut
  {NoStop}%
\bibitem [{\citenamefont {Mai}\ \emph {et~al.}(2022{\natexlab{b}})\citenamefont
  {Mai}, \citenamefont {Zhao}, \citenamefont {Feldman},\ and\ \citenamefont
  {Phillips}}]{arxiv2}%
  \BibitemOpen
  \bibfield  {author} {\bibinfo {author} {\bibfnamefont {P.}~\bibnamefont
  {Mai}}, \bibinfo {author} {\bibfnamefont {J.}~\bibnamefont {Zhao}}, \bibinfo
  {author} {\bibfnamefont {B.~E.}\ \bibnamefont {Feldman}},\ and\ \bibinfo
  {author} {\bibfnamefont {P.~W.}\ \bibnamefont {Phillips}},\ }\bibfield
  {title} {\bibinfo {title} {1/4 is the new 1/2: Interaction-induced quantum
  anomalous and spin \mbox{Hall} \mbox{Mott} insulators},\ }\href@noop {} {\
  (\bibinfo {year} {2022}{\natexlab{b}})},\ \Eprint
  {https://arxiv.org/abs/arXiv:2210.11486} {arXiv:2210.11486} \BibitemShut
  {NoStop}%
\bibitem [{\citenamefont {Hatsugai}\ and\ \citenamefont {Kohmoto}(1992)}]{HK}%
  \BibitemOpen
  \bibfield  {author} {\bibinfo {author} {\bibfnamefont {Y.}~\bibnamefont
  {Hatsugai}}\ and\ \bibinfo {author} {\bibfnamefont {M.}~\bibnamefont
  {Kohmoto}},\ }\bibfield  {title} {\bibinfo {title} {Exactly solvable model of
  correlated lattice electrons in any dimensions},\ }\href
  {https://doi.org/10.1143/JPSJ.61.2056} {\bibfield  {journal} {\bibinfo
  {journal} {Journal of the Physical Society of Japan}\ }\textbf {\bibinfo
  {volume} {61}},\ \bibinfo {pages} {2056} (\bibinfo {year} {1992})},\ \Eprint
  {https://arxiv.org/abs/https://doi.org/10.1143/JPSJ.61.2056}
  {https://doi.org/10.1143/JPSJ.61.2056} \BibitemShut {NoStop}%
\bibitem [{\citenamefont {Byczuk}\ and\ \citenamefont
  {Spa\l{}ek}(1994{\natexlab{a}})}]{Byczuk1994}%
  \BibitemOpen
  \bibfield  {author} {\bibinfo {author} {\bibfnamefont {K.}~\bibnamefont
  {Byczuk}}\ and\ \bibinfo {author} {\bibfnamefont {J.}~\bibnamefont
  {Spa\l{}ek}},\ }\bibfield  {title} {\bibinfo {title} {Statistical properties
  and statistical interaction for particles with spin: The \mbox{Hubbard} model
  in one dimension and a statistical spin liquid},\ }\href
  {https://doi.org/10.1103/PhysRevB.50.11403} {\bibfield  {journal} {\bibinfo
  {journal} {Phys. Rev. B}\ }\textbf {\bibinfo {volume} {50}},\ \bibinfo
  {pages} {11403} (\bibinfo {year} {1994}{\natexlab{a}})}\BibitemShut {NoStop}%
\bibitem [{\citenamefont {Byczuk}\ and\ \citenamefont
  {Spa\l{}ek}(1995)}]{Byczuk1995}%
  \BibitemOpen
  \bibfield  {author} {\bibinfo {author} {\bibfnamefont {K.}~\bibnamefont
  {Byczuk}}\ and\ \bibinfo {author} {\bibfnamefont {J.}~\bibnamefont
  {Spa\l{}ek}},\ }\bibfield  {title} {\bibinfo {title} {Universality classes,
  statistical exclusion principle, and properties of interacting fermions},\
  }\href {https://doi.org/10.1103/PhysRevB.51.7934} {\bibfield  {journal}
  {\bibinfo  {journal} {Phys. Rev. B}\ }\textbf {\bibinfo {volume} {51}},\
  \bibinfo {pages} {7934} (\bibinfo {year} {1995})}\BibitemShut {NoStop}%
\bibitem [{\citenamefont {Byczuk}\ and\ \citenamefont
  {Spa\l{}ek}(1994{\natexlab{b}})}]{Spalek1994}%
  \BibitemOpen
  \bibfield  {author} {\bibinfo {author} {\bibfnamefont {K.}~\bibnamefont
  {Byczuk}}\ and\ \bibinfo {author} {\bibfnamefont {J.}~\bibnamefont
  {Spa\l{}ek}},\ }\bibfield  {title} {\bibinfo {title} {Application of
  statistical spin liquid concept to high temperature superconductivity},\
  }\href {http://przyrbwn.icm.edu.pl/APP/PDF/85/a085z2p20.pdf} {\bibfield
  {journal} {\bibinfo  {journal} {Acta Physica Polonica A}\ }\textbf {\bibinfo
  {volume} {85}},\ \bibinfo {pages} {337} (\bibinfo {year}
  {1994}{\natexlab{b}})}\BibitemShut {NoStop}%
\bibitem [{\citenamefont {Huang}\ \emph {et~al.}(2022)\citenamefont {Huang},
  \citenamefont {Nave},\ and\ \citenamefont {Phillips}}]{Nature1}%
  \BibitemOpen
  \bibfield  {author} {\bibinfo {author} {\bibfnamefont {E.~W.}\ \bibnamefont
  {Huang}}, \bibinfo {author} {\bibfnamefont {G.~L.}\ \bibnamefont {Nave}},\
  and\ \bibinfo {author} {\bibfnamefont {P.~W.}\ \bibnamefont {Phillips}},\
  }\bibfield  {title} {\bibinfo {title} {Discrete symmetry breaking defines the
  \mbox{Mott} quartic fixed point},\ }\href
  {https://doi.org/10.1038/s41567-022-01529-8} {\bibfield  {journal} {\bibinfo
  {journal} {Nature Physics}\ }\textbf {\bibinfo {volume} {18}},\ \bibinfo
  {pages} {511} (\bibinfo {year} {2022})}\BibitemShut {NoStop}%
\bibitem [{\citenamefont {Phillips}\ \emph {et~al.}(2020)\citenamefont
  {Phillips}, \citenamefont {Yeo},\ and\ \citenamefont {Huang}}]{Nature2}%
  \BibitemOpen
  \bibfield  {author} {\bibinfo {author} {\bibfnamefont {P.~W.}\ \bibnamefont
  {Phillips}}, \bibinfo {author} {\bibfnamefont {L.}~\bibnamefont {Yeo}},\ and\
  \bibinfo {author} {\bibfnamefont {E.~W.}\ \bibnamefont {Huang}},\ }\bibfield
  {title} {\bibinfo {title} {Exact theory for superconductivity in a doped
  \mbox{Mott} insulator},\ }\href {https://doi.org/10.1038/s41567-020-0988-4}
  {\bibfield  {journal} {\bibinfo  {journal} {Nature Physics}\ }\textbf
  {\bibinfo {volume} {16}},\ \bibinfo {pages} {1175} (\bibinfo {year}
  {2020})}\BibitemShut {NoStop}%
\bibitem [{\citenamefont {Chang}\ \emph {et~al.}(2013)\citenamefont {Chang},
  \citenamefont {Zhang}, \citenamefont {Feng}, \citenamefont {Shen},
  \citenamefont {Zhang}, \citenamefont {Guo}, \citenamefont {Li}, \citenamefont
  {Ou}, \citenamefont {Wei}, \citenamefont {Wang}, \citenamefont {Ji},
  \citenamefont {Feng}, \citenamefont {Ji}, \citenamefont {Chen}, \citenamefont
  {Jia}, \citenamefont {Dai}, \citenamefont {Fang}, \citenamefont {Zhang},
  \citenamefont {He}, \citenamefont {Wang}, \citenamefont {Lu}, \citenamefont
  {Ma},\ and\ \citenamefont {Xue}}]{Chang2013}%
  \BibitemOpen
  \bibfield  {author} {\bibinfo {author} {\bibfnamefont {C.-Z.}\ \bibnamefont
  {Chang}}, \bibinfo {author} {\bibfnamefont {J.}~\bibnamefont {Zhang}},
  \bibinfo {author} {\bibfnamefont {X.}~\bibnamefont {Feng}}, \bibinfo {author}
  {\bibfnamefont {J.}~\bibnamefont {Shen}}, \bibinfo {author} {\bibfnamefont
  {Z.}~\bibnamefont {Zhang}}, \bibinfo {author} {\bibfnamefont
  {M.}~\bibnamefont {Guo}}, \bibinfo {author} {\bibfnamefont {K.}~\bibnamefont
  {Li}}, \bibinfo {author} {\bibfnamefont {Y.}~\bibnamefont {Ou}}, \bibinfo
  {author} {\bibfnamefont {P.}~\bibnamefont {Wei}}, \bibinfo {author}
  {\bibfnamefont {L.-L.}\ \bibnamefont {Wang}}, \bibinfo {author}
  {\bibfnamefont {Z.-Q.}\ \bibnamefont {Ji}}, \bibinfo {author} {\bibfnamefont
  {Y.}~\bibnamefont {Feng}}, \bibinfo {author} {\bibfnamefont {S.}~\bibnamefont
  {Ji}}, \bibinfo {author} {\bibfnamefont {X.}~\bibnamefont {Chen}}, \bibinfo
  {author} {\bibfnamefont {J.}~\bibnamefont {Jia}}, \bibinfo {author}
  {\bibfnamefont {X.}~\bibnamefont {Dai}}, \bibinfo {author} {\bibfnamefont
  {Z.}~\bibnamefont {Fang}}, \bibinfo {author} {\bibfnamefont {S.-C.}\
  \bibnamefont {Zhang}}, \bibinfo {author} {\bibfnamefont {K.}~\bibnamefont
  {He}}, \bibinfo {author} {\bibfnamefont {Y.}~\bibnamefont {Wang}}, \bibinfo
  {author} {\bibfnamefont {L.}~\bibnamefont {Lu}}, \bibinfo {author}
  {\bibfnamefont {X.-C.}\ \bibnamefont {Ma}},\ and\ \bibinfo {author}
  {\bibfnamefont {Q.-K.}\ \bibnamefont {Xue}},\ }\bibfield  {title} {\bibinfo
  {title} {Experimental observation of the quantum anomalous \mbox{Hall} effect
  in a magnetic topological insulator},\ }\href
  {https://doi.org/10.1126/science.1234414} {\bibfield  {journal} {\bibinfo
  {journal} {Science}\ }\textbf {\bibinfo {volume} {340}},\ \bibinfo {pages}
  {167} (\bibinfo {year} {2013})},\ \Eprint
  {https://arxiv.org/abs/https://www.science.org/doi/pdf/10.1126/science.1234414}
  {https://www.science.org/doi/pdf/10.1126/science.1234414} \BibitemShut
  {NoStop}%
\bibitem [{\citenamefont {Chang}\ \emph {et~al.}(2015)\citenamefont {Chang},
  \citenamefont {Zhao}, \citenamefont {Kim}, \citenamefont {Zhang},
  \citenamefont {Assaf}, \citenamefont {Heiman}, \citenamefont {Zhang},
  \citenamefont {Liu}, \citenamefont {Chan},\ and\ \citenamefont
  {Moodera}}]{Chang2015}%
  \BibitemOpen
  \bibfield  {author} {\bibinfo {author} {\bibfnamefont {C.-Z.}\ \bibnamefont
  {Chang}}, \bibinfo {author} {\bibfnamefont {W.}~\bibnamefont {Zhao}},
  \bibinfo {author} {\bibfnamefont {D.~Y.}\ \bibnamefont {Kim}}, \bibinfo
  {author} {\bibfnamefont {H.}~\bibnamefont {Zhang}}, \bibinfo {author}
  {\bibfnamefont {B.~A.}\ \bibnamefont {Assaf}}, \bibinfo {author}
  {\bibfnamefont {D.}~\bibnamefont {Heiman}}, \bibinfo {author} {\bibfnamefont
  {S.-C.}\ \bibnamefont {Zhang}}, \bibinfo {author} {\bibfnamefont
  {C.}~\bibnamefont {Liu}}, \bibinfo {author} {\bibfnamefont {M.~H.~W.}\
  \bibnamefont {Chan}},\ and\ \bibinfo {author} {\bibfnamefont {J.~S.}\
  \bibnamefont {Moodera}},\ }\bibfield  {title} {\bibinfo {title}
  {High-precision realization of robust quantum anomalous \mbox{Hall} state in
  a hard ferromagnetic topological insulator},\ }\href
  {https://doi.org/10.1038/nmat4204} {\bibfield  {journal} {\bibinfo  {journal}
  {Nature Materials}\ }\textbf {\bibinfo {volume} {14}},\ \bibinfo {pages}
  {473} (\bibinfo {year} {2015})}\BibitemShut {NoStop}%
\bibitem [{\citenamefont {Sharpe}\ \emph {et~al.}(2019)\citenamefont {Sharpe},
  \citenamefont {Fox}, \citenamefont {Barnard}, \citenamefont {Finney},
  \citenamefont {Watanabe}, \citenamefont {Taniguchi}, \citenamefont
  {Kastner},\ and\ \citenamefont {Goldhaber-Gordon}}]{Sharpe2019}%
  \BibitemOpen
  \bibfield  {author} {\bibinfo {author} {\bibfnamefont {A.~L.}\ \bibnamefont
  {Sharpe}}, \bibinfo {author} {\bibfnamefont {E.~J.}\ \bibnamefont {Fox}},
  \bibinfo {author} {\bibfnamefont {A.~W.}\ \bibnamefont {Barnard}}, \bibinfo
  {author} {\bibfnamefont {J.}~\bibnamefont {Finney}}, \bibinfo {author}
  {\bibfnamefont {K.}~\bibnamefont {Watanabe}}, \bibinfo {author}
  {\bibfnamefont {T.}~\bibnamefont {Taniguchi}}, \bibinfo {author}
  {\bibfnamefont {M.~A.}\ \bibnamefont {Kastner}},\ and\ \bibinfo {author}
  {\bibfnamefont {D.}~\bibnamefont {Goldhaber-Gordon}},\ }\bibfield  {title}
  {\bibinfo {title} {Emergent ferromagnetism near three-quarters filling in
  twisted bilayer graphene},\ }\href {https://doi.org/10.1126/science.aaw3780}
  {\bibfield  {journal} {\bibinfo  {journal} {Science}\ }\textbf {\bibinfo
  {volume} {365}},\ \bibinfo {pages} {605} (\bibinfo {year} {2019})},\ \Eprint
  {https://arxiv.org/abs/https://www.science.org/doi/pdf/10.1126/science.aaw3780}
  {https://www.science.org/doi/pdf/10.1126/science.aaw3780} \BibitemShut
  {NoStop}%
\bibitem [{\citenamefont {Deng}\ \emph {et~al.}(2020)\citenamefont {Deng},
  \citenamefont {Yu}, \citenamefont {Shi}, \citenamefont {Guo}, \citenamefont
  {Xu}, \citenamefont {Wang}, \citenamefont {Chen},\ and\ \citenamefont
  {Zhang}}]{Deng2020}%
  \BibitemOpen
  \bibfield  {author} {\bibinfo {author} {\bibfnamefont {Y.}~\bibnamefont
  {Deng}}, \bibinfo {author} {\bibfnamefont {Y.}~\bibnamefont {Yu}}, \bibinfo
  {author} {\bibfnamefont {M.~Z.}\ \bibnamefont {Shi}}, \bibinfo {author}
  {\bibfnamefont {Z.}~\bibnamefont {Guo}}, \bibinfo {author} {\bibfnamefont
  {Z.}~\bibnamefont {Xu}}, \bibinfo {author} {\bibfnamefont {J.}~\bibnamefont
  {Wang}}, \bibinfo {author} {\bibfnamefont {X.~H.}\ \bibnamefont {Chen}},\
  and\ \bibinfo {author} {\bibfnamefont {Y.}~\bibnamefont {Zhang}},\ }\bibfield
   {title} {\bibinfo {title} {Quantum anomalous \mbox{Hall} effect in intrinsic
  magnetic topological insulator mnbi<sub>2</sub>te<sub>4</sub>},\ }\href
  {https://doi.org/10.1126/science.aax8156} {\bibfield  {journal} {\bibinfo
  {journal} {Science}\ }\textbf {\bibinfo {volume} {367}},\ \bibinfo {pages}
  {895} (\bibinfo {year} {2020})},\ \Eprint
  {https://arxiv.org/abs/https://www.science.org/doi/pdf/10.1126/science.aax8156}
  {https://www.science.org/doi/pdf/10.1126/science.aax8156} \BibitemShut
  {NoStop}%
\bibitem [{\citenamefont {Serlin}\ \emph {et~al.}(2020)\citenamefont {Serlin},
  \citenamefont {Tschirhart}, \citenamefont {Polshyn}, \citenamefont {Zhang},
  \citenamefont {Zhu}, \citenamefont {Watanabe}, \citenamefont {Taniguchi},
  \citenamefont {Balents},\ and\ \citenamefont {Young}}]{Serlin2020}%
  \BibitemOpen
  \bibfield  {author} {\bibinfo {author} {\bibfnamefont {M.}~\bibnamefont
  {Serlin}}, \bibinfo {author} {\bibfnamefont {C.~L.}\ \bibnamefont
  {Tschirhart}}, \bibinfo {author} {\bibfnamefont {H.}~\bibnamefont {Polshyn}},
  \bibinfo {author} {\bibfnamefont {Y.}~\bibnamefont {Zhang}}, \bibinfo
  {author} {\bibfnamefont {J.}~\bibnamefont {Zhu}}, \bibinfo {author}
  {\bibfnamefont {K.}~\bibnamefont {Watanabe}}, \bibinfo {author}
  {\bibfnamefont {T.}~\bibnamefont {Taniguchi}}, \bibinfo {author}
  {\bibfnamefont {L.}~\bibnamefont {Balents}},\ and\ \bibinfo {author}
  {\bibfnamefont {A.~F.}\ \bibnamefont {Young}},\ }\bibfield  {title} {\bibinfo
  {title} {Intrinsic quantized anomalous \mbox{Hall} effect in a moir\&\#xe9;
  heterostructure},\ }\href {https://doi.org/10.1126/science.aay5533}
  {\bibfield  {journal} {\bibinfo  {journal} {Science}\ }\textbf {\bibinfo
  {volume} {367}},\ \bibinfo {pages} {900} (\bibinfo {year} {2020})},\ \Eprint
  {https://arxiv.org/abs/https://www.science.org/doi/pdf/10.1126/science.aay5533}
  {https://www.science.org/doi/pdf/10.1126/science.aay5533} \BibitemShut
  {NoStop}%
\bibitem [{\citenamefont {Satake}\ \emph {et~al.}(2020)\citenamefont {Satake},
  \citenamefont {Shiogai}, \citenamefont {Mazur}, \citenamefont {Kimura},
  \citenamefont {Awaji}, \citenamefont {Fujiwara}, \citenamefont {Nojima},
  \citenamefont {Nomura}, \citenamefont {Souma}, \citenamefont {Sato},
  \citenamefont {Dietl},\ and\ \citenamefont {Tsukazaki}}]{Satake2020}%
  \BibitemOpen
  \bibfield  {author} {\bibinfo {author} {\bibfnamefont {Y.}~\bibnamefont
  {Satake}}, \bibinfo {author} {\bibfnamefont {J.}~\bibnamefont {Shiogai}},
  \bibinfo {author} {\bibfnamefont {G.~P.}\ \bibnamefont {Mazur}}, \bibinfo
  {author} {\bibfnamefont {S.}~\bibnamefont {Kimura}}, \bibinfo {author}
  {\bibfnamefont {S.}~\bibnamefont {Awaji}}, \bibinfo {author} {\bibfnamefont
  {K.}~\bibnamefont {Fujiwara}}, \bibinfo {author} {\bibfnamefont
  {T.}~\bibnamefont {Nojima}}, \bibinfo {author} {\bibfnamefont
  {K.}~\bibnamefont {Nomura}}, \bibinfo {author} {\bibfnamefont
  {S.}~\bibnamefont {Souma}}, \bibinfo {author} {\bibfnamefont
  {T.}~\bibnamefont {Sato}}, \bibinfo {author} {\bibfnamefont {T.}~\bibnamefont
  {Dietl}},\ and\ \bibinfo {author} {\bibfnamefont {A.}~\bibnamefont
  {Tsukazaki}},\ }\bibfield  {title} {\bibinfo {title} {Magnetic-field-induced
  topological phase transition in \mbox{Fe}-doped \mbox{(Bi,Sb)$_2$Se$_3$}
  heterostructures},\ }\href
  {https://doi.org/10.1103/PhysRevMaterials.4.044202} {\bibfield  {journal}
  {\bibinfo  {journal} {Phys. Rev. Mater.}\ }\textbf {\bibinfo {volume} {4}},\
  \bibinfo {pages} {044202} (\bibinfo {year} {2020})}\BibitemShut {NoStop}%
\bibitem [{\citenamefont {Fijalkowski}\ \emph {et~al.}(2020)\citenamefont
  {Fijalkowski}, \citenamefont {Hartl}, \citenamefont {Winnerlein},
  \citenamefont {Mandal}, \citenamefont {Schreyeck}, \citenamefont {Brunner},
  \citenamefont {Gould},\ and\ \citenamefont {Molenkamp}}]{Fijalkowski2020}%
  \BibitemOpen
  \bibfield  {author} {\bibinfo {author} {\bibfnamefont {K.~M.}\ \bibnamefont
  {Fijalkowski}}, \bibinfo {author} {\bibfnamefont {M.}~\bibnamefont {Hartl}},
  \bibinfo {author} {\bibfnamefont {M.}~\bibnamefont {Winnerlein}}, \bibinfo
  {author} {\bibfnamefont {P.}~\bibnamefont {Mandal}}, \bibinfo {author}
  {\bibfnamefont {S.}~\bibnamefont {Schreyeck}}, \bibinfo {author}
  {\bibfnamefont {K.}~\bibnamefont {Brunner}}, \bibinfo {author} {\bibfnamefont
  {C.}~\bibnamefont {Gould}},\ and\ \bibinfo {author} {\bibfnamefont {L.~W.}\
  \bibnamefont {Molenkamp}},\ }\bibfield  {title} {\bibinfo {title}
  {Coexistence of surface and bulk ferromagnetism mimics skyrmion \mbox{Hall}
  effect in a topological insulator},\ }\href
  {https://doi.org/10.1103/PhysRevX.10.011012} {\bibfield  {journal} {\bibinfo
  {journal} {Phys. Rev. X}\ }\textbf {\bibinfo {volume} {10}},\ \bibinfo
  {pages} {011012} (\bibinfo {year} {2020})}\BibitemShut {NoStop}%
\bibitem [{\citenamefont {Pournaghavi}\ \emph {et~al.}(2021)\citenamefont
  {Pournaghavi}, \citenamefont {Islam}, \citenamefont {Islam}, \citenamefont
  {Autieri}, \citenamefont {Dietl},\ and\ \citenamefont
  {Canali}}]{Pournaghavi2021}%
  \BibitemOpen
  \bibfield  {author} {\bibinfo {author} {\bibfnamefont {N.}~\bibnamefont
  {Pournaghavi}}, \bibinfo {author} {\bibfnamefont {M.~F.}\ \bibnamefont
  {Islam}}, \bibinfo {author} {\bibfnamefont {R.}~\bibnamefont {Islam}},
  \bibinfo {author} {\bibfnamefont {C.}~\bibnamefont {Autieri}}, \bibinfo
  {author} {\bibfnamefont {T.}~\bibnamefont {Dietl}},\ and\ \bibinfo {author}
  {\bibfnamefont {C.~M.}\ \bibnamefont {Canali}},\ }\bibfield  {title}
  {\bibinfo {title} {Realization of the \mbox{Chern}-insulator and
  axion-insulator phases in antiferromagnetic
  \mbox{MnTe/Bi$_2$(Se,Te)$_3$/MnTe} heterostructures},\ }\href
  {https://doi.org/10.1103/PhysRevB.103.195308} {\bibfield  {journal} {\bibinfo
   {journal} {Phys. Rev. B}\ }\textbf {\bibinfo {volume} {103}},\ \bibinfo
  {pages} {195308} (\bibinfo {year} {2021})}\BibitemShut {NoStop}%
\bibitem [{\citenamefont {Li}\ \emph {et~al.}(2021)\citenamefont {Li},
  \citenamefont {Jiang}, \citenamefont {Shen}, \citenamefont {Zhang},
  \citenamefont {Li}, \citenamefont {Tao}, \citenamefont {Devakul},
  \citenamefont {Watanabe}, \citenamefont {Taniguchi}, \citenamefont {Fu},
  \citenamefont {Shan},\ and\ \citenamefont {Mak}}]{Li2021}%
  \BibitemOpen
  \bibfield  {author} {\bibinfo {author} {\bibfnamefont {T.}~\bibnamefont
  {Li}}, \bibinfo {author} {\bibfnamefont {S.}~\bibnamefont {Jiang}}, \bibinfo
  {author} {\bibfnamefont {B.}~\bibnamefont {Shen}}, \bibinfo {author}
  {\bibfnamefont {Y.}~\bibnamefont {Zhang}}, \bibinfo {author} {\bibfnamefont
  {L.}~\bibnamefont {Li}}, \bibinfo {author} {\bibfnamefont {Z.}~\bibnamefont
  {Tao}}, \bibinfo {author} {\bibfnamefont {T.}~\bibnamefont {Devakul}},
  \bibinfo {author} {\bibfnamefont {K.}~\bibnamefont {Watanabe}}, \bibinfo
  {author} {\bibfnamefont {T.}~\bibnamefont {Taniguchi}}, \bibinfo {author}
  {\bibfnamefont {L.}~\bibnamefont {Fu}}, \bibinfo {author} {\bibfnamefont
  {J.}~\bibnamefont {Shan}},\ and\ \bibinfo {author} {\bibfnamefont {K.~F.}\
  \bibnamefont {Mak}},\ }\bibfield  {title} {\bibinfo {title} {Quantum
  anomalous \mbox{Hall} effect from intertwined moir\'e bands},\ }\href
  {https://doi.org/10.1038/s41586-021-04171-1} {\bibfield  {journal} {\bibinfo
  {journal} {Nature}\ }\textbf {\bibinfo {volume} {600}},\ \bibinfo {pages}
  {641} (\bibinfo {year} {2021})}\BibitemShut {NoStop}%
\bibitem [{\citenamefont {Hussain}\ \emph {et~al.}(2023)\citenamefont
  {Hussain}, \citenamefont {Fakhredine}, \citenamefont {Islam}, \citenamefont
  {Sattigeri}, \citenamefont {Autieri},\ and\ \citenamefont
  {Cuono}}]{Hussain2023}%
  \BibitemOpen
  \bibfield  {author} {\bibinfo {author} {\bibfnamefont {G.}~\bibnamefont
  {Hussain}}, \bibinfo {author} {\bibfnamefont {A.}~\bibnamefont {Fakhredine}},
  \bibinfo {author} {\bibfnamefont {R.}~\bibnamefont {Islam}}, \bibinfo
  {author} {\bibfnamefont {R.~M.}\ \bibnamefont {Sattigeri}}, \bibinfo {author}
  {\bibfnamefont {C.}~\bibnamefont {Autieri}},\ and\ \bibinfo {author}
  {\bibfnamefont {G.}~\bibnamefont {Cuono}},\ }\bibfield  {title} {\bibinfo
  {title} {Correlation-driven topological transition in \mbox{Janus}
  two-dimensional vanadates},\ }\bibfield  {journal} {\bibinfo  {journal}
  {Materials}\ }\textbf {\bibinfo {volume} {16}},\ \href
  {https://doi.org/10.3390/ma16041649} {10.3390/ma16041649} (\bibinfo {year}
  {2023})\BibitemShut {NoStop}%
\bibitem [{\citenamefont {Chang}\ \emph {et~al.}(2023)\citenamefont {Chang},
  \citenamefont {Liu},\ and\ \citenamefont {MacDonald}}]{Chang2023}%
  \BibitemOpen
  \bibfield  {author} {\bibinfo {author} {\bibfnamefont {C.-Z.}\ \bibnamefont
  {Chang}}, \bibinfo {author} {\bibfnamefont {C.-X.}\ \bibnamefont {Liu}},\
  and\ \bibinfo {author} {\bibfnamefont {A.~H.}\ \bibnamefont {MacDonald}},\
  }\bibfield  {title} {\bibinfo {title} {Colloquium: Quantum anomalous
  \mbox{Hall} effect},\ }\href {https://doi.org/10.1103/RevModPhys.95.011002}
  {\bibfield  {journal} {\bibinfo  {journal} {Rev. Mod. Phys.}\ }\textbf
  {\bibinfo {volume} {95}},\ \bibinfo {pages} {011002} (\bibinfo {year}
  {2023})}\BibitemShut {NoStop}%
\bibitem [{\citenamefont {Meng}\ \emph {et~al.}(2016)\citenamefont {Meng},
  \citenamefont {Huang}, \citenamefont {Peng}, \citenamefont {Li},
  \citenamefont {Chen}, \citenamefont {Xu}, \citenamefont {Zhang},
  \citenamefont {Wang},\ and\ \citenamefont {Zhang}}]{Meng2016}%
  \BibitemOpen
  \bibfield  {author} {\bibinfo {author} {\bibfnamefont {Z.}~\bibnamefont
  {Meng}}, \bibinfo {author} {\bibfnamefont {L.}~\bibnamefont {Huang}},
  \bibinfo {author} {\bibfnamefont {P.}~\bibnamefont {Peng}}, \bibinfo {author}
  {\bibfnamefont {D.}~\bibnamefont {Li}}, \bibinfo {author} {\bibfnamefont
  {L.}~\bibnamefont {Chen}}, \bibinfo {author} {\bibfnamefont {Y.}~\bibnamefont
  {Xu}}, \bibinfo {author} {\bibfnamefont {C.}~\bibnamefont {Zhang}}, \bibinfo
  {author} {\bibfnamefont {P.}~\bibnamefont {Wang}},\ and\ \bibinfo {author}
  {\bibfnamefont {J.}~\bibnamefont {Zhang}},\ }\bibfield  {title} {\bibinfo
  {title} {Experimental observation of a topological band gap opening in
  ultracold \mbox{Fermi} gases with two-dimensional spin-orbit coupling},\
  }\href {https://doi.org/10.1103/PhysRevLett.117.235304} {\bibfield  {journal}
  {\bibinfo  {journal} {Phys. Rev. Lett.}\ }\textbf {\bibinfo {volume} {117}},\
  \bibinfo {pages} {235304} (\bibinfo {year} {2016})}\BibitemShut {NoStop}%
\bibitem [{\citenamefont {Huang}\ \emph {et~al.}(2016)\citenamefont {Huang},
  \citenamefont {Meng}, \citenamefont {Wang}, \citenamefont {Peng},
  \citenamefont {Zhang}, \citenamefont {Chen}, \citenamefont {Li},
  \citenamefont {Zhou},\ and\ \citenamefont {Zhang}}]{Huang2016}%
  \BibitemOpen
  \bibfield  {author} {\bibinfo {author} {\bibfnamefont {L.}~\bibnamefont
  {Huang}}, \bibinfo {author} {\bibfnamefont {Z.}~\bibnamefont {Meng}},
  \bibinfo {author} {\bibfnamefont {P.}~\bibnamefont {Wang}}, \bibinfo {author}
  {\bibfnamefont {P.}~\bibnamefont {Peng}}, \bibinfo {author} {\bibfnamefont
  {S.-L.}\ \bibnamefont {Zhang}}, \bibinfo {author} {\bibfnamefont
  {L.}~\bibnamefont {Chen}}, \bibinfo {author} {\bibfnamefont {D.}~\bibnamefont
  {Li}}, \bibinfo {author} {\bibfnamefont {Q.}~\bibnamefont {Zhou}},\ and\
  \bibinfo {author} {\bibfnamefont {J.}~\bibnamefont {Zhang}},\ }\bibfield
  {title} {\bibinfo {title} {Experimental realization of two-dimensional
  synthetic spin–orbit coupling in ultracold \mbox{Fermi} gases},\ }\href
  {https://doi.org/10.1038/nphys3672} {\bibfield  {journal} {\bibinfo
  {journal} {Nature Physics}\ }\textbf {\bibinfo {volume} {12}},\ \bibinfo
  {pages} {540} (\bibinfo {year} {2016})}\BibitemShut {NoStop}%
\bibitem [{\citenamefont {Wang}\ \emph {et~al.}(2012)\citenamefont {Wang},
  \citenamefont {Qi},\ and\ \citenamefont {Zhang}}]{Zhang_2012}%
  \BibitemOpen
  \bibfield  {author} {\bibinfo {author} {\bibfnamefont {Z.}~\bibnamefont
  {Wang}}, \bibinfo {author} {\bibfnamefont {X.-L.}\ \bibnamefont {Qi}},\ and\
  \bibinfo {author} {\bibfnamefont {S.-C.}\ \bibnamefont {Zhang}},\ }\bibfield
  {title} {\bibinfo {title} {Topological invariants for interacting topological
  insulators with inversion symmetry},\ }\href
  {https://doi.org/10.1103/PhysRevB.85.165126} {\bibfield  {journal} {\bibinfo
  {journal} {Phys. Rev. B}\ }\textbf {\bibinfo {volume} {85}},\ \bibinfo
  {pages} {165126} (\bibinfo {year} {2012})}\BibitemShut {NoStop}%
\bibitem [{\citenamefont {Wang}\ and\ \citenamefont
  {Zhang}(2012)}]{Zhang_2012_2}%
  \BibitemOpen
  \bibfield  {author} {\bibinfo {author} {\bibfnamefont {Z.}~\bibnamefont
  {Wang}}\ and\ \bibinfo {author} {\bibfnamefont {S.-C.}\ \bibnamefont
  {Zhang}},\ }\bibfield  {title} {\bibinfo {title} {Simplified topological
  invariants for interacting insulators},\ }\href
  {https://doi.org/10.1103/PhysRevX.2.031008} {\bibfield  {journal} {\bibinfo
  {journal} {Phys. Rev. X}\ }\textbf {\bibinfo {volume} {2}},\ \bibinfo {pages}
  {031008} (\bibinfo {year} {2012})}\BibitemShut {NoStop}%
\bibitem [{\citenamefont {Niu}\ \emph {et~al.}(1985)\citenamefont {Niu},
  \citenamefont {Thouless},\ and\ \citenamefont {Wu}}]{Thouless1985}%
  \BibitemOpen
  \bibfield  {author} {\bibinfo {author} {\bibfnamefont {Q.}~\bibnamefont
  {Niu}}, \bibinfo {author} {\bibfnamefont {D.~J.}\ \bibnamefont {Thouless}},\
  and\ \bibinfo {author} {\bibfnamefont {Y.-S.}\ \bibnamefont {Wu}},\
  }\bibfield  {title} {\bibinfo {title} {Quantized hall conductance as a
  topological invariant},\ }\href {https://doi.org/10.1103/PhysRevB.31.3372}
  {\bibfield  {journal} {\bibinfo  {journal} {Phys. Rev. B}\ }\textbf {\bibinfo
  {volume} {31}},\ \bibinfo {pages} {3372} (\bibinfo {year}
  {1985})}\BibitemShut {NoStop}%
\bibitem [{\citenamefont {Fukui}\ \emph {et~al.}(2005)\citenamefont {Fukui},
  \citenamefont {Hatsugai},\ and\ \citenamefont {Suzuki}}]{Fukui_2005}%
  \BibitemOpen
  \bibfield  {author} {\bibinfo {author} {\bibfnamefont {T.}~\bibnamefont
  {Fukui}}, \bibinfo {author} {\bibfnamefont {Y.}~\bibnamefont {Hatsugai}},\
  and\ \bibinfo {author} {\bibfnamefont {H.}~\bibnamefont {Suzuki}},\
  }\bibfield  {title} {\bibinfo {title} {Chern numbers in discretized
  \mbox{Brillouin} zone: Efficient method of computing (spin) \mbox{Hall}
  conductances},\ }\href {https://doi.org/10.1143/JPSJ.74.1674} {\bibfield
  {journal} {\bibinfo  {journal} {Journal of the Physical Society of Japan}\
  }\textbf {\bibinfo {volume} {74}},\ \bibinfo {pages} {1674} (\bibinfo {year}
  {2005})},\ \Eprint
  {https://arxiv.org/abs/https://doi.org/10.1143/JPSJ.74.1674}
  {https://doi.org/10.1143/JPSJ.74.1674} \BibitemShut {NoStop}%
\end{thebibliography}

%

\end{document}